\documentclass[12pt]{amsart}
\usepackage[mathscr]{eucal}
\usepackage{amsmath,amsfonts,amssymb,amsthm}
\usepackage{hyperref}
\usepackage[matrix,arrow]{xy}
\usepackage{times}
\def\Fedosov{Fedosov:1994}
\def\AKSZ{Alexandrov:1997kv}
\def\BGST{Barnich:2004cr}
\newcommand{\BGadS}{Barnich:2006pc}
\newcommand{\BGunf}{Barnich:2005ru}

\def\BG{Barnich:2003wj}
\def\BGL{Batalin:2001je}

\def\BGLns{Batalin:2005df}
\advance\textheight\topskip
\renewcommand{\tilde}{\widetilde}
\renewcommand{\hat}{\widehat}
\newtheorem{prop}{Proposition}[section]

\newtheorem{rem}{Remark}[section]

\def\os{{\rm on-shell}}
\def\sd{S^\dagger}
\def\bsd{\bar S^\dagger}

\newcommand{\T}{\mathrm{T}}

\newcommand{\bref}[1]{\textbf{\ref{#1}}}

\newcommand{\dmn}{d}  

\newcommand{\im}{\mathop{\mathrm{Im}}}

\newcommand{\p}[1]{|#1|}
\newcommand{\gh}[1]{\mathrm{gh}(#1)}

\newcommand{\assalgebra}{\mathscr}    
\newcommand{\algA}{\assalgA}
\newcommand{\algH}{\mathfrak{h}}
\newcommand{\assalgA}{\assalgebra{A}}

\newcommand{\algg}{\Liealg{g}}

\newcommand{\Liealg}{\mathfrak}       

\newcommand{\map}{\,\mathrm{:}\,}

\newcommand{\dd}{\partial}
\renewcommand{\d}{\partial}

\newcommand{\tensor}{\otimes}

\renewcommand{\geq}{\,{\geqslant}\,}

\newcommand{\binner}[2]{%
  {\langle}\kern-4.15pt{\langle}#1{,}\,#2{\rangle}\kern-4.15pt{\rangle}}
\newcommand{\commut}[2]{[#1{,}\,#2]}
\newcommand{\qcommut}[2]{[#1{,}\,#2]_*}
\newcommand{\scommut}[2]{\{#1{,}\,#2\}_*}
\newcommand{\pb}[2]{\left\{{}#1{},{}#2{}\right\}}

\newcommand{\half}{\mathchoice{%
    \ffrac{1}{2}}{\frac{1}{2}}{\frac{1}{2}}{\frac{1}{2}}}

\newcommand{\hhalf}{\ffrac{1}{2}}

\newcommand{\ffrac}[2]{\raisebox{.5pt}%
  {\footnotesize$\displaystyle\frac{#1}{#2}$}\kern1pt}

\newcommand{\brst}{\mathsf{\Omega}}

\newcommand{\dl}[1]{\mathchoice{\ffrac{\dd}{\dd #1}}{\frac{\dd}{\dd
      #1}}{\ffrac{\dd}{\dd #1}}{\ffrac{\dd}{\dd #1}}}

\newcommand{\ddr}[2]{\ffrac{\dd_R #1}{\dd #2}}

\newcommand{\vdl}[1]{\ffrac{{\delta}}{\delta #1}}

\def\rank{\mathop\mathrm{rank}\nolimits}

\newcommand{\Pj}{\mathscr{{P}}}

\newcommand{\cd}{{c^\dagger}}

\newcommand{\bundle}{\boldsymbol}

\newcommand{\derham}{\boldsymbol{d}}

\newcommand{\manifold}[1]{\mathscr{#1}}
\newcommand{\manX}{\manifold{X}}

\newcommand{\manM}{\manifold{M}}

\def\cC{\mathcal{C}}

\def\cE{\mathcal{E}}
\def\cF{\mathcal{F}}

\def\cH{\mathcal{H}}

\def\cP{\mathcal{P}}

\def\bA{\boldsymbol{A}}
\def\bB{\boldsymbol{B}}

\def\bF{\boldsymbol{F}}

\def\bR{\boldsymbol{R}}
\numberwithin{equation}{section} \makeatletter
\@addtoreset{equation}{section}

\def\@secnumfont{\bfseries}
\def\subsubsection{\@startsection{subsubsection}{3}%
  \z@{.5\linespacing\@plus.7\linespacing}{-.5em}%
  {\normalfont\bfseries}}
\def\paragraph{\@startsection{paragraph}{4}%
  \z@\z@{-\fontdimen2\font}%
  \normalfont\bfseries}
\def\subparagraph{\@startsection{subparagraph}{5}%
  \z@\z@{-\fontdimen2\font}%
  \normalfont\bfseries}

\makeatother

\begin{document}
\def\mytitle{Off-shell Gauge Fields from BRST Quantization}

\def\shorttitle{\mytitle}

\pagestyle{myheadings}
\markboth{\textsc{\small\mytitle}}{%
  \textsc{\small\mytitle}}
\addtolength{\headsep}{4pt}

\begin{flushright}\small
ULB-TH/06-09\\
FIAN-TD/06-06\\
\texttt{hep-th/0605089}\\[-3pt]
\end{flushright}

\begin{centering}

  \vspace{1cm}

  \textbf{\Large{\mytitle}}

\vspace{2cm}

  {\large Maxim Grigoriev$^{a,b}$ }

\vspace{1cm}

\begin{minipage}{.9\textwidth}\small \it \begin{center}
   $^a$Tamm Theory Department, Lebedev Physics
   Institute,\\ Leninsky prospect 53, 119991 Moscow, Russia\end{center}
\end{minipage}

    \vspace{.5cm}
\begin{minipage}{.9\textwidth}\small \it \begin{center}
   $^b$Physique Th\'eorique et Math\'ematique, Universit\'e Libre de
   Bruxelles\\ and \\ International Solvay Institutes, \\ Campus
   Plaine C.P. 231, B-1050 Bruxelles, Belgium \end{center}
\end{minipage}

\end{centering}

\vspace{1cm}

\begin{center}
  \begin{minipage}{.9\textwidth}
    \textsc{Abstract}.  We propose a construction for nonlinear
    off-shell gauge field theories based on a constrained system
    quantized in the sense of deformation quantization.  The key idea
    is to consider the star-product BFV--BRST master equation as an
    equation of motion.  The construction is formulated in terms of
    the BRST extention of the unfolded formalism that can also be
    understood as an appropriate generalization of the AKSZ procedure.
    As an application, we consider a very simple constrained system, a
    quantized scalar particle, and show that it gives rise to an
    off-shell higher-spin gauge theory that automatically appears in
    the parent form and properly takes the familiar trace constraint
    into account.  In particular, we derive a geometrically
    transparent form of the off-shell higher-spin theory on the AdS
    background.
  \end{minipage}
\end{center}

\thispagestyle{empty} \newpage

\begin{small}
  \tableofcontents
\end{small}

\section{Introduction}

An off-shell field theory is by definition a theory whose equations of
motion are equivalent to algebraic constraints. This implies that one
can in principle solve the equations of motion and eliminate pure
gauge degrees of freedom in order to obtain the unconstrained fields
and gauge symmetries.  Although the off-shell formulation does not
actually describe dynamical equations, it can be useful from various
standpoints.

In particular, an off-shell theory can encode all the information on
the field content and gauge symmetries in the form adapted for
introducing consistent interactions.  For example, this applies to the
higher-spin (HS) theories, where the relevance of the off-shell
formulation in constructing nonlinear equations~\cite{Vasiliev:2003ev}
(see also \cite{Bekaert:2005vh} for a review) has been recently
realized in~\cite{Sagnotti:2005ns}.  More recently, a compact and
geometrically transparent form of the nonlinear off-shell HS theory on
Minkowski space was constructed by M.~Vasiliev
in~\cite{Vasiliev:2005zu}.

An off-shell theory can also be regarded as an intermediate step in
constructing a Lagrangian formulation. Indeed, being algebraic, the
off-shell equations of motion can always be made Lagrangian by
introducing Lagrange multipliers (we refer the reader
to~\cite{Vasiliev:2005zu} for a more extensive discussion).  Moreover,
under some regularity assumptions, any off-shell theory can be
equivalently formulated in the Lagrangian Batalin--Vilkovisky
formalism~\cite{Batalin:1981jr,Batalin:1983jr} by introducing the so-called generalized Lagrange
multipliers~\cite{\BGadS}.

In this paper, we propose the generating procedure for constructing
off-shell gauge theories starting from a (quantized) constrained
system.  The underlying idea is to identify the $*$-product version of
the quantum Batalin--Fradkin--Vilkovisky
(BFV)~\cite{Fradkin:1975cq,Batalin:1977pb} master equation
$\ffrac{1}{2\hbar}\qcommut{\brst}{\brst}=0$ for a constrained
Hamiltonian system as a dynamical equation for some gauge field
theory.  Namely, fields are identified as coefficients in the
expansion of $\brst$ with respect to ghost variables and possibly some
extra variables present in the formalism.

The idea to interpret constraints as dynamical fields does not seem
completely new.  Similar approaches have been developed, e.g.,
in~\cite{Bars:2000qm,Segal:2000ke,Segal:2002gd,Vasiliev:2005zu,Sagnotti:2005ns,Engquist:2005yt}
without using the BRST technique but by analyzing the consistency of
the constraint algebra for the particular constrained systems.
Somewhat analogous ideas underly the Lagrangian considerations
in~\cite{Hata:1993pp}.  The general construction developed in this
paper in terms of BRST theory provides a unified framework that allows
considering more general constrained systems such that the symmetries
present in the model are manifest.

In more technical terms, we use the BRST extention~\cite{\BGunf} of
the Vasiliev unfolded
formalism~\cite{Vasiliev:1990en,Vasiliev:1992av,Vasiliev:2003ev} that
can also be understood as the non-Lagrangian version of the
Alexandrov--Kontsevich--Schwartz--Zaboronsky (AKSZ) sigma
model~\cite{\AKSZ} with the target space being the quantum (in the
sense of a $*$-product) extended phase space of a constrained system.
Equations of motion for such a system can be identified as a component
form of the BFV quantum master equation for an appropriately extended
constrained system.  The extension is the one used in the Fedosov
quantization~\cite{\Fedosov,Fedosov-book} (or its version adapted to
the case of cotangent bundles~\cite{Bordemann:1997er}) and generalized
to the case of systems with
constraints~\cite{Batalin:1989mb,\BGL,\BGLns} (see also
\cite{\BGST,\BGadS} for the particular constrained systems relevant in
the present context).

Expanding the theory around a given solution of the master equation
gives, in particular, the linearized theory that can be naturally
identified as a parent form~\cite{\BGST,\BGadS} of the field theory
associated with some BRST first-quantized model.  In the two examples
explicitly considered in the paper, this BRST first-quantized system
is the off-shell version of a parent system for HS fields on the
flat~\cite{\BGST} and AdS space~\cite{\BGadS} respectively.

\bigskip

The paper is organized as follows. In Section~\bref{sec:basic}, we
propose an elementary construction for off-shell gauge theories from
quantum constrained systems and discuss their linearizations around
particular solutions.  The simplest example of a system with just one
constraint is then explicitly considered.

In Section~\bref{sec:2}, we review the BRST extension of the nonlinear
unfolded formalism and consistent reductions in this framework,
originally described in~\cite{\BGunf}.  The linearization of such
theories is then briefly discussed.  The general construction for
off-shell gauge theories starting from quantum constrained systems is
given in Section~\bref{sec:general} within the BRST-extended unfolded
formulation.

Section~\bref{sec:HS} is devoted to explicit examples leading to the
off-shell HS gauge theories on various backgrounds.  In
Section~\bref{sec:ads}, we propose a compact form of the off-shell HS
theory on AdS space in terms of the embedding space and discuss its
relation to the Vasiliev unfolded formulation.

\section{Master equation as an equation of motion}
\label{sec:basic}
\subsection{Off-shell gauge theories from quantum constrained systems}
\label{sec:basic-1}

An interesting class of gauge theories can be obtained starting from a
quantum phase space of a quantum constrained system. Suppose we are
given with a constrained system quantized in the sense of deformation
quantization. This implies the associative $*$-product algebra $\algA$
depending on the quantization parameter $\hbar$. In what follows we
restrict ourselves to the formal deformation quantization and
therefore allow $\algA$ to be the algebra of formal power series in
$\hbar$ with the coefficients being functions on the extended phase
space of the system (i.e.  functions in phase space coordinates and
ghost variables).  The associative $*$-product on $\algA$ reduces to the
pointwise multiplication and the Poisson bracket in the $\hbar\to 0$
limit:
\begin{equation}
  f*g=fg+O(\hbar)\,,\qquad f*g-(-1)^{\p{f}\p{g}}g*f
=\hbar\pb{f}{g}+O(\hbar^2)\,,
\end{equation}
where $\pb{\cdot}{\cdot}$ denotes the Poisson bracket on the extended
phase space and $\p{\cdot}$ denotes the Grassmann parity. $\algA$ is also equipped with the ghost number grading denoted by
$\gh{\cdot}$.  For simplicity we suppose that no physical fermions are
present so that $\p{\phi}=\gh{\phi}\mod 2$ for any homogeneous
$\phi\in\algA$.

Let us assume in addition that the extended phase space of the system
is a bundle over a manifold $\manX_0$ identified as the space-time
manifold in what follows.  Algebra $\algA$ can be then considered as
that of sections of the appropriate associated vector bundle
$\bundle\cH$ over $\manX_0$ with the fiber being the linear space
$\cH$ of functions on the fiber (here and below we denote by
$\bundle\cH$ the vector bundle with the fiber isomorphic to a vector
space $\cH$). In what follows we assume $\cH$ to be (graded)
finite-dimensional, i.e. that one can always find a suitable degree
such that each homogeneous component is
finite-dimensional.~\footnote{We do not give a precise definitions
  here.  However, in the concrete systems discussed in what follows
  $\cH$ is usually an algebra of polynomials or formal power series so
  that it suitably decomposes into the finite-dimensional subspaces of
  elements with fixed homogeneity.}

Let $e_A$ be a basis in $\cH$.  Any element of $\algA$ can be (locally
on $\manX_0$) represented as $\chi=e_A \chi^A(x)$ where $x^\mu$ are
local coordinates on $\manX_0$.  In this representation the
$*$-product is a bilinear bidifferential operation on sections of
$\bundle\cH$. In addition we assume that $\qcommut{x^\mu}{x^\nu}=0$
where $x^\mu,x^\nu$ are considered as elements of $\algA$.  Note that
relaxing this condition corresponds to the interesting possibility to
describe noncommutative gauge field theories. To simplify the
exposition we also assume all the bundles to be trivial unless
otherwise specified or, which is the same, restrict ourselves to the
local analysis. In particular, $\algA$ can be then identified with
the $\cH$-valued functions on $\manX_0$

A quantum BRST charge (more precisely, a symbol of the BRST operator) is an
element $\Phi=e_A\Phi^A(x)\in\algA$ satisfying the master equation
along with the ghost number and the Grassmann parity assignments:
\begin{equation}
\label{eq:master}
  \ffrac{1}{2\hbar}\qcommut{\Phi}{\Phi}=0\,,  \qquad \gh{\Phi}=1\,,\quad  \p{\Phi}=1\,.
\end{equation}
Note that it follows from $\gh{\Phi}=1$ that $\Phi^A=0$ for $e_A$ with
$\gh{e_A}\neq 1$. Under the standard regularity assumptions on the
constraints the quantum BRST charge properly describes the quantum constrained
system, at least at the level of deformation quantization (i.e. the
algebra of quantum observables).  The specification of the correct
space of quantum states equipped with the inner product is a separate
question which we do not discuss here.

Instead of considering the master equation as a generating equation
for the constraint algebra of the quantum constrained system we are
going to interpret it in terms of some classical gauge field theory.
The key idea is to consider components $\Phi^A(x)$ as dynamical fields
defined on the space-time manifold $\manX_0$ and the master equation
as the equation of motion for $\Phi^A$. The field theory defined in
this way is invariant under the natural gauge symmetries determined by
the master equation itself:
\begin{equation}
\label{eq:gs}
  \delta_\Lambda \Phi=\ffrac{1}{\hbar}\qcommut{\Phi}{\Lambda}\,,
\end{equation}
where gauge parameter $\Lambda$ is an arbitrary element from $\algA$
with $\gh{\Lambda}=0$ and $\p{\Lambda}=0$.

The physical interpretation of the constructed gauge field theory has
to do with the off-shell description of the background fields. The
constraints of the system can be identified with (the generating
functions for) the background fields determining e.g. configuration
space geometry, background Maxwell field etc.  The master equation
ensuring the consistency of the constraints imposes the equations
(usually equivalent to algebraic i.e. not containing derivatives with
respect to $x^\mu$) on the background fields and determines their
gauge symmetries.  In Section~\bref{sec:basic-example} we consider the
scalar-particle system (see also the discussion
in~\cite{Segal:2001di,Segal:2002gd}) where this interpretation is
transparent.

Suppose $\Phi_0$ be a particular solution to~\eqref{eq:master}.  The
equations of motion and gauge transformations expanded around $\Phi_0$
read as
\begin{equation}
\label{eq:exp}
  \ffrac{1}{\hbar}\qcommut{\Phi_0}{\Phi}
+\ffrac{1}{2\hbar}\qcommut{\Phi}{\Phi}=0\,,\qquad 
\delta_\Lambda \Phi=\ffrac{1}{\hbar}\qcommut{\Phi_0}{\Lambda}
+\ffrac{1}{\hbar}\qcommut{\Phi}{\Lambda}\,,
\end{equation}
where by slight abusing notations the fluctuation around $\Phi_0$ is
again denoted by $\Phi$. The terms linear in $\Phi$ determines the
linearized equations of motion and gauge symmetries.

The linearized theory can be naturally interpreted as a field theory
associated to a BRST first-quantized system
$(\brst,\Gamma(\bundle\cH,\manX_0))$ with the ``space of states''
being $\Gamma(\bundle\cH,\manX_0)\cong \algA$ and nilpotent BRST
operator $\brst$ defined by
$\brst\phi=\hbar^{-1}\qcommut{\Phi_0}{\phi}$ for any $\phi\in \algA$.
Here and in what follows we use notation
$(\brst,\Gamma(\bundle\cH,\manX_0))$ for the first-quantized BRST
system specified by the ``space of states''
$\Gamma(\bundle\cH,\manX_0)$ (the space of sections of the vector
bundle $\bundle\cH$ over $\manX_0$) and the BRST operator
$\brst\map\Gamma(\bundle\cH,\manX_0)\to\Gamma(\bundle\cH,\manX_0)$.
 Equations \eqref{eq:exp} can be then rewritten as
\begin{equation}
\label{eq:exp-2}
  \brst\Phi+\ffrac{1}{2\hbar}\qcommut{\Phi}{\Phi}=0\,,\qquad 
\delta_\Lambda \Phi=\brst{\Lambda}+\ffrac{1}{\hbar}\qcommut{\Phi}{\Lambda}\,,
\end{equation}
so that their linear parts indeed take the familiar form $\brst\Phi=0$
and $\delta_\Lambda\Phi=\brst\Lambda$
(see~\cite{\BG,\BGST,Gaberdiel:1997ia} and references therein for more
details on field theories associated to the first-quantized BRST
systems).

Suppose $\brst$ be an odd nilpotent operator $\brst\map\algA\to\algA$
not necessarily generated by some $\Phi_0$. In this case
equations~\eqref{eq:exp-2} still determines a consistent gauge field
theory provided $\gh{\brst}=1$ and $\brst$ satisfies Leinbitz rule
$\brst (\phi*\chi)
=(\brst\phi)*\chi+(-1)^{\p{\phi}\p{\chi}}\phi*(\brst\chi)$. This
possibility along with further generalizations of the construction are
discussed in more details in Section~\bref{sec:diff-s}.

Note that in our approach the space of states appears to be the space
$\algA$ of functions on the extended phase space. It does not therefore coincide with the
space of quantum states (if it was specified) of the starting point
constrained system because the later can be (at least formally)
identified with the suitable functions on the Lagrangian submanifold
of the phase space (i.e.  functions in only one half of the phase
space coordinates). In order to describe quantum states in terms of
$\algA$ one needs additional factorization procedure which we do not
discuss in this paper.

Let us also note that instead of the $*$-product description one can
use the standard language of operators. Moreover, all the
constructions can be also reformulated in these terms provided one is
given with the suitable representation space.  In particular, if
variables $x^\mu$ are quantized in the coordinate representation the
operators can be identified as differential operators on $\manX_0$
with coefficients in linear operators on the ``internal'' space of
states. Such a representation for a constraint operator of a scalar
particle has been used in~\cite{Segal:2002gd} in the related context.

All the considerations of this section remaines true if one takes the
classical limit by replacing $\algA$ with the commutative algebra of
the phase space functions equipped with the Poisson bracket (i.e. the
classical limit of the $*$-product algebra).  This can also be
understood as replacing the quantum constrained system with the
classical one (its classical limit).

\subsection{The basic example}\label{sec:basic-example}

To illustrate the construction let us consider nearly the simplest
constrained system, a ``scalar particle'' on the flat Minkowski space
$\manX_0$, with only one constraint. Let $p_\mu$ be the momenta
conjugated to coordinates $x^\mu$ on $\manX_0$ and $F(x,p)$ the
constraint. In order to handle the constraint in the BRST approach we
introduce Grassmann odd ghost variables $c,\pi$ with
$\gh{c}=1,\gh{\pi}=-1$ and $\qcommut{c}{\pi}=-\hbar$. Variables
$x^\mu,p_\nu$ are assumed to carry vanishing ghost degree. The quantum
BRST charge is then given by
\begin{equation}
\Phi=c F(x,p)  
\end{equation}
and automatically satisfies $\qcommut{\Phi}{\Phi}=0$ because $F$ is the only
constraint present in the model.

According to the general strategy the dynamical fields are
coefficients in the expansion of $F$ with respect to momenta $p_\mu$
\begin{equation}
  F(x,p)=\phi_0(x)+\phi_1^\mu(x)p_\mu+\phi_2^{\mu\nu}p_\mu p_\nu+\phi_3^{\mu\nu\rho}(x)+\ldots\,.
\end{equation}
and can be identified with the symmetric tensor fields on $\manX_0$.
The gauge symmetries are determined by~\eqref{eq:gs} where in this
case $\Lambda=\lambda(x;p)+c\pi \chi(x;p)$ (note that terms with
nonzero ghost structure do not enter as one can easily see by counting the ghost
degree). Explicitly one gets
\begin{equation}
  \delta_\Lambda F=\ffrac{1}{\hbar}\qcommut{F}{\lambda}+\half\scommut{F}{\chi}\,.
\end{equation}
where $\scommut{a}{b}=a*b+(-1)^{\p{a}\p{b}}b*a$ denotes the graded
symmetric $*$-anticommutator.

Let us expand the theory around a particular solution
$\Phi_0=c\,\hhalf\eta^{\mu\nu}p_\mu p_\nu$ describing a scalar particle on
Minkowski space. In this case the gauge transformations take the form
\begin{equation}
\label{eq:gs-particle}
\delta_\Lambda F= -p^\mu \dl{x^\mu}\lambda+\hhalf p^2\chi+\ffrac{\hbar^2}{8}\dl{x^\mu}\dl{x^\mu}\chi+
\ffrac{1}{\hbar}\qcommut{F}{\lambda}+\hhalf\scommut{F}{\chi}\,,
\end{equation}
where the first three terms give a linearized gauge transformation.
Note that this gauge symmetry has been originally considered
in~\cite{Segal:2002gd} from a slightly different perspective.

In order to identify the off-shell gauge field theory described by
$\Phi=c(\hhalf p^2+F(x;p))$ let us concentrate on the linearized
theory.  Using the linearized gauge transformation given by the first
three terms in~\eqref{eq:gs-particle} one can always achieve
$\dl{p_\mu}\dl{p^\mu}F=0$, i.e., the tensor fields entering $F(x;p)$
can be assumed traceless. This restricts the gauge transformations to
those with $\chi=0$ and modifies the remaining gauge transformations
by the appropriate projector to the traceless component.

There are two interpretations of the resulting off-shell theory.  The
traceless symmetric tensor fields subjected to the gauge
transformation above provides the off-shell definition of the
conformal HS theory~\cite{Fradkin:1985am,Fradkin:1989md}. It then
follows that the off-shell theory determined by~\eqref{eq:gs-particle}
can be also considered as an off-shell description of the conformal HS
fields.  We will not discuss conformal HS theory and refer instead
to~\cite{Segal:2002gd}, where, in particular, conformal HS theory was
constructed in the analogous terms. Although in this case we only
reproduced the description from~\cite{Segal:2002gd} the advantage of
our approach is that it can be uniformly extended to more general
quantum constrained systems.

Another interpretation of the off-shell theory just constructed has to
do with the Fronsdal HS gauge
theory~\cite{Fronsdal:1978rb,Fronsdal:1979vb}.  Namely, we show (see
Section~\bref{sec:flat}) that the off-shell theory determined
by~\eqref{eq:gs-particle} is equivalent through the elimination of
generalized auxiliary fields to the off-shell theory for the Fronsdal
HS fields in the parent form~\cite{\BGST} (see
section~\bref{sec:reduction} for definition of generalized auxiliary
fields). More precisely, in~\bref{sec:flat} we construct the
appropriate extention of the model~\eqref{eq:gs-particle}, which
provides a geometrically transparent formulation of the (off-shell)
Fronsdal HS theory.  Note that the extended model can also serve as an
off-shell theory for conformal HS fields.

To directly see the relation with the conventional formulation of the
Fronsdal theory suppose that the following equations have been in
addition imposed on $F$:
\begin{equation}
\label{eq:part-contr}
  \dl{x^\mu}\dl{x_\mu}F(x;p)=0\,,\qquad   \dl{x^\mu}\dl{p_\mu}F(x;p)=0\,.
\end{equation}
Together with the condition $\dl{p_\mu}\dl{p^\mu}F=0$ this coincides
with the equations of motion and the partial gauge fixing conditions
for Fronsdal HS fields identified in~\cite{Mikhailov:2002bp}.  It is
important to note, however, that this does not imply that the theory
determined by~\eqref{eq:part-contr} along with
$\dl{p_\mu}\dl{p^\mu}F=0$ and the remaining gauge symmetry is
equivalent to the on-shell Fronsdal HS theory in the strong sense
(i.e. through the elimination/addition of generalized auxiliary fields).

\subsection{The field theory BRST differential}\label{sec:diff-s}
It is useful to reformulate the procedure in the BRST theory terms.
Here, we closely follow the non-Lagrangian BRST formulation
from~\cite{\BGST} (see also~\cite{\BGunf,\BGadS} and references
therein). Let $e_A$ be a basis in $\cH$.  We then associate a
supermanifold $\manM$ to the superspace $\cH$.  To this end we assign
a variable $\psi^A$ to each basis element $e_A$ and prescribe
$\gh{\psi_A}=1-\gh{e_A}$, $\p{\psi_A}=1+\p{e_A} \mod 2$. One then
defines $\manM$ to be a supermanifold\footnote{Note that one can
  either take the real basis in $\cH$ or the complex one. Independent
  fields $\psi^A$ are then to be taken real or complex accordingly. In
  any case $\manM$ is the complex supermanifold described either in
  terms of complex coordinates or the real coordinates and the complex
  structure. In order to end up with e.g.  Fronsdal model with the
  real fields one needs to impose in addition an appropriate reality
  condition.  We do not discuss reality conditions in the general
  setting assuming that we are working with the complexified versions
  of the respective theories. Note, however, that in the examples
  considered in the paper the required reality conditions are rather
  obvious (see e.g.~\cite{\BGST} for the case of Fronsdal HS theory).}
with coordinates $\psi^A$. In order to define $\manM$ one also needs
to fix the class of functions in $\psi^A$. Although most of our
present considerations do not really depend on this choice, for
definiteness we take smooth functions.  In what follows we call
$\manM$ the supermanifold associated to $\cH$.

Consider the field theory with fields $\psi^A$ defined on the
space-time manifold $\manX_0$. The interpretation of $\psi^A$ depends
on its ghost number. In particular, physical fields are those with
vanishing ghost number. If $\gh{\psi^A}\neq 0$ then $\psi^A$ should be
considered as a ghost field or an antifield. The BRST differential
determining the theory is given by
\begin{equation}
\label{eq:s-def}
  s\psi^A=\ffrac{1}{2\hbar}\qcommut{e_B}{e_C}^A\psi^C\psi^B(-1)^{\p{B}}\,,
\end{equation}
where $\p{B}=\p{\psi^B}$ and $\qcommut{\phi}{\chi}^A$ denotes the
$e_A$ component of $\qcommut{\phi}{\chi}$, i.e.,
$\qcommut{\phi}{\chi}=\qcommut{\phi}{\chi}^A e_A$. In what follows we
mainly utilize the jet-space formulation of local gauge field
theories.  In this formulation fields $\psi^A$ and their space-time
derivatives are treated as independent coordinates on the jet space.
The BRST differential is the vector field on the jet space determined
by~\eqref{eq:s-def} and the condition $\commut{s}{\d_\mu}=0$ where
$\d_\mu$ is a total derivative (see e.g.~\cite{Barnich:2000zw} for
more details).

It is useful to introduce the so-called string field $\Psi=e_A \tensor
\psi^A$ which is understood as an element of $\cH \tensor
\cC^\infty({\manM})$ (in what follows we simply write
$\Psi=e_A\psi^A$, see~\cite{\BGST} for more details). In terms of the
string field the definition of the BRST differential $s$ takes the
form
\begin{equation}
 s\Psi=\ffrac{1}{2\hbar}\qcommut{\Psi}{\Psi}\,.
\end{equation}
It is also useful to expand $\Psi$ into components containing fields
at given ghost degree: $\Psi=\sum_k \Psi^{(k)}$ where
$\Psi^{(k)}=e_{A_k}\psi^{A_k}$ with $\gh{\psi^{A_k}}=k$.  Note that
contrary to the conventional string field associated with the space of
states of the first-quantized system, $\Psi$ is associated to the
algebra of functions on the entire extended phase space. However, as
are going to see, $\Psi$ can be naturally interpreted as a
conventional string field (but associated with the different quantum
constrained system) if one considers the linearized theory.

The BRST differential determines the equations of motion, the gauge
transformation, and the reducibility conditions along with higher
order structures of the gauge algebra. In particular, if $\psi^{A_k}$
denote component fields entering $\Psi^{(k)}$ (i.e.
$\gh{\psi^{A_k}}=k$) then equations of motion and the gauge
transformations have the form
\begin{equation}
\label{eq:em-gs-gen}
(s\psi^{A_{-1}})\big|_{\psi^{A_k}=0,\,k\neq 0}=0\,,\qquad
\delta\psi^{A_0}=(s\psi^{A_0})\big|_{{\psi^{A_k}}=0,\,k\neq 0,1}\,,
\end{equation}
with ghost-number-$1$ fields $\psi^{A_1}$ replaced by gauge parameters
$\lambda^{A_1}$ with $\p{\lambda^{A_1}}=\p{\psi^{A_1}}+1\, \mod 2$.

Note that if the theory does not contain physical fermionic fields (as
we have assumed) then all ghost-number-zero fields $\psi^{A_0}$ are
bosonic and can be identified with coefficients $\Phi^{A_0}$ in the
expansion of $\Phi$ with respect to the basis $e_{A_0}$ of the
ghost-number-one subspace in $\cH$. Analogously, the gauge parameters
correspond to ghost-number-one fields $\psi^{A_1}$ associated to the
basis elements of the ghost-number-zero subspace. However, if one wants
to consider theories containing physical fermionic fields or build the
complete BV-BRST description one needs to replace the coefficients
$\Phi^A(x)$ in the expansion of a generic element from $\algA$ with
the fields $\psi^A$ and to prescribe $\gh{\psi^A}=1-\gh{e_A}$ and
$\p{\psi^A}=1-\p{e_A} \mod 2$, i.e. to replace superspace $\cH$ with
the supermanifold $\manM$.

The theory expanded around a particular solution to the equations of
motion can also be compactly formulated in the BV-BRST terms. Namely,
let $\Psi_0$ be a particular solution to
$\derham\Psi_0+(2\hbar)^{-1}\qcommut{\Psi_0}{\Psi_0}=0$. The BRST
differential of the theory expanded around $\Psi_0$ is then given by
\begin{equation}
\label{eq:sss}
  s\Psi=\brst\Psi+\ffrac{1}{2\hbar}\qcommut{\Psi}{\Psi}\,,
\end{equation}
where BRST operator $\brst$ is defined by
$\brst\phi=\hbar^{-1}\qcommut{\Psi_0}{\phi}$ for $\phi\in \algA$.

As we have already seen this linearized theory can be considered as
the BRST field theory associated to the first-quantized system
$(\brst,\Gamma(\bundle{\cH},\manX_0))$. Indeed, the linearized BRST
differential has the familiar form $s_0\Psi=\brst\Psi$. From this
point of view $\Psi$ is to be identified with the string field
associated with the space of states $\Gamma(\bundle{\cH},\manX_0)$.
It is in this sense $\Psi$ is related to the conventional string field
used in the context of string field theory (see
e.g.~\cite{Gaberdiel:1997ia}).  To be precise one also needs to adjust
the ghost number grading in $\algA$ in order to fit the standard
convention $\gh{\Psi}=0$.

The construction of this section can be naturally extended to a more
general class of quantum constrained system. To see this let us
consider the theory expanded around a particular solution to the
master equation $\Psi_0$ which induces the map
$\brst\map\algA\to\algA$ by $\brst\phi=\qcommut{\Psi_0}{\phi}$.
Together with the bilinear map induced by the $*$-commutator
$\qcommut{\cdot}{\cdot}\map\algA\times \algA \to \algA$ these two maps
define the differential graded Lie algebra structure on $\algA$. Even
if $\brst$ can not be represented as $\qcommut{\Phi_0}{\cdot}$ for some
$\Psi_0$ one can still define a consistent gauge theory determined
by~\eqref{eq:exp-2} or, equivalently, by the BRST
differential~\eqref{eq:sss}. In particular, for the BRST field theory
we construct in Section~\bref{sec:general} the differential
$s\Psi=\brst\Psi+(2\hbar)^{-1}\qcommut{\Psi}{\Psi}$ is precisely of
this type, i.e. $\brst$ can not be represented as
$\brst=\qcommut{\Psi_0}{\cdot}$.

More generally, one can replace the differential graded Lie algebra
with the more general structure known as $L_\infty$
algebra~\cite{Lada:1993wc} which is specified by a collection of
polylinear graded-antisymmetric operations:
\begin{equation}
\label{eq:maps}
  \brst\map \algA\to \algA\,,\qquad \qcommut{\cdot}{\cdot}\map \algA\times\algA \to \algA\,,\qquad
\qcommut{\cdot}{\cdot}^{3}\map \algA\times\algA\times\algA \to \algA\,,\qquad \ldots
\end{equation}
that satisfy certain compatibility conditions generalizing those of the
differential graded Lie algebra. In this case one can still consider
the consistent gauge field theory determined by the following BRST differential:
\begin{equation}
\label{eq:L-infty-s}
  s\Psi=\brst\Psi+\ffrac{1}{2\hbar}\qcommut{\Psi}{\Psi}+\ffrac{1}{6\hbar^2}[\Psi,\Psi,\Psi]^3+\ldots\,.
\end{equation}
Note that the nilpotency of $s$ is equivalent to the defining
relations of $L_\infty$-algebra. Moreover, $L_\infty$-algebra
structure is usually defined in terms of an odd nilpotent odd vector
field on the associated supermanifold.

Equation~\eqref{eq:L-infty-s} generalizes the construction to the case
of quantum constrained systems described by the $A_\infty$
algebra~\cite{Stasheff:1963}. In this case the $A_\infty$-structure is
determined by the nilpotent $\brst\map\algA\to\algA$, the homotopy
associative $*$-product (i.e.  associative only in the
$\brst$-cohomology), and the higher order polylinear maps.  The
$L_\infty$-structure is then obtained from the $A_\infty$-structure by
taking the graded antisymmetrization of the polylinear maps. Quantum
constrained systems of this type naturally arise in quantization of
some classical constrained systems (see
e.g.~\cite{Lyakhovich:2004xd,\BGLns}).

\section{Off-shell gauge theories in the BRST extended unfolded formulation}
\label{sec:2}

\subsection{BRST extension of the non-linear unfolded formalism}
In this section we briefly recall the BRST extension~\cite{\BGunf} of
the Vasiliev non-linear unfolded
formalism~\cite{Vasiliev:1990en,Vasiliev:1992av,Vasiliev:2003ev},
proposed recently by G.~Barnich and the present author.

Consider two supermanifolds: a supermanifold $\manX$ and $\manM$
playing the roles of (appropriately extended) space-time and the
target space manifolds respectively.  Let $\manX$ be a supermanifold
equipped with a degree ${\mathrm{gh}}_\manX({\cdot\,})$, an odd
nilpotent vector field $\derham,\,\mathrm{gh}_\manX(\derham)=1$, and a
volume form $d\mu$ preserved by $\derham$. Let the supermanifold
$\manM$ be equipped with a degree $\mathrm{gh}_\manM$, an odd
nilpotent vector field $Q$, $\mathrm{gh}_\manM(Q)=1$. In the
literature a supermanifold equipped with an odd nilpotent vector field
is often called $Q$-manifold while the vector field itself is referred
to as $Q$-structure~\cite{Schwarz:1992gs}.

The basic example for $\manX$ is the odd tangent bundle $\Pi T\manX_0$
which has a natural volume form and is equipped with the De Rham
differential $\derham$.  If $x^\mu$ are local coordinates on $\manX_0$
and $\theta^\mu$ are associated coordinates on the fibers of $\Pi
T\manX_0$ then $\derham=\theta^\mu \dl{x^\mu}$ and one assumes in
addition $\mathrm{gh}_\manX(x^\mu)=0$ and
$\mathrm{gh}_\manX(\theta^\mu)=1$. Functions on $\manX$ are then
identified with the differential forms on $\manX_0$ via
$dx^\mu=\theta^\mu$ while the degree $\mathrm{gh}_\manX$ is just a
standard form degree.

Consider the manifold of smooth maps from $\manX$ to $\manM$. This space is
naturally equipped with the total degree denoted by $\gh{\cdot}$ and
an odd nilpotent vector field $s$, $\gh s=1$. Indeed, it is a standard
geometrical fact that any vector field on the ``space-time'' manifold
or the ``target space'' manifold determines a vector field on the
space of maps. More precisely, if $z^\alpha$ are local coordinates on $\manX$
(in the case where $\manX=\Pi T\manX_0$ coordinates $z^\alpha$ split into
$x^\mu$ and $\theta^\mu$) and $\psi^A$ are local coordinates on
$\manM$, the expression for $s$ reads
\begin{equation}
\label{eq:s-AKSZ}
  s=\int_{\manX}d\mu (-1)^{\p{d\mu}}\left[
\derham\psi^A(z)+Q^A(\psi(z))
\right]
\vdl{\psi^A(z)}\,.
\end{equation}
Vector field $s$ can be considered as a BRST differential of a field
theory on $\manX$. Indeed, the basic properties $s^2=0$ and $\gh{s}=1$
hold.  In what follows we refer to this system as a quadruple
$(\manX,\derham,\manM,Q)$.

For the system $(\manX,\derham,\manM,Q)$ it is easy to check using the
explicit form~\eqref{eq:s-AKSZ} that
\begin{equation}
\label{eq:jet-s}
s\psi^A=\derham\psi^A+Q^A(\psi)\,.
\end{equation}
This equation can be taken as a definition of the BRST differential in
the jet-bundle description of the theory. In this approach component
fields $(\psi_p)^A_{\mu_1\ldots\mu_p}$ entering
$\psi^A(x,\theta)=(\psi_0)^A(x)+\theta^\mu(\psi_1)^A_\mu(x)+\ldots$
and their derivatives with resoect to $x^\mu$ are treated as
independent coordinates on the jet space.

Let $\psi^{A_k}$ denote component fields with $\gh{\psi^{A_k}}=k$.
Using the explicit form~\eqref{eq:jet-s} of the BRST differential one
finds the component form of the equations of motion and gauge symmetries
\begin{equation}
\left(\derham\psi^{A}+Q^{A}(\psi)\right)\big|_{\psi^{A_k}=0,\,k\neq 0}=0\,,
\end{equation}
and
\begin{equation}
\delta_\lambda \psi^{A}=
\left(\derham \psi^{A}+Q^{A}(\psi)\right)
\big|_{\psi^{A_1}=\lambda^{A_1},\,\psi^{A_k}=0,\,k\neq 0,1}\,,
\end{equation}
where $\delta_\lambda\psi^A$ denotes variation of $\psi^A$ under the
gauge variation of its physical component fields $\psi^{A_0}$. In
particular, if $\gh{\psi^A}\geq 0$ for all fields then the equations
of motion determine the so-called free differential
algebra~\cite{Sullivan}. If one does not require $\gh{\psi^A}\geq 0$
then the equations of motion can also contain some constraints.

In general, instead of $(\manX,\derham,\manM,Q)$ one can similarly
consider a fibered bundle with $\manX$ being a base manifold, a fiber
isomorphic to $\manM$, and the transition functions preserving
$Q$-structure. In this case the field space is the space of sections
of the bundle instead of $\manM$-valued functions.  However, for the
sake of simplicity we do not consider here non-trivial bundles unless
otherwise specified. Note that the general construction anyway reduces
to $(\manX,\derham,\manM,Q)$ locally.

In the case where the ``target'' supermanifold $\manM$ is in addition
equipped with a compatible Poisson bracket (antibracket)
$\pb{\cdot}{\cdot}$ and $Q=\pb{S}{\cdot}$ is generated by a ``master
action'' (``BRST charge'') $S$ satisfying the classical master
equation $\pb{S}{S}=0$, one can construct a field theory master action
$\mathbf{S}$ on the space of maps. This procedure was proposed
in~\cite{Alexandrov:1997kv} as an approach for constructing BV-BRST
formulations of topological sigma models
(see~\cite{Cattaneo:1999fm,Batalin:2001fh,Cattaneo:2001ys,Park:2000au,Roytenberg:2002nu,\BG,
Lyakhovich:2004xd,Kazinski:2005eb,Ikeda:2006wd}
and references therein for further developments and applications).  A
generalization that also includes the Hamiltonian BRST formulation has
been proposed in~\cite{Grigoriev:1999qz} and covers the case where
$\mathbf{S}$ is Grassmann odd and is to be interpreted as a BRST
charge of the BFV-BRST formulation of the theory.

A simplest but characteristic example is provided by taking $\manM$
to be $\Pi \algg$ with $\algg$ being a Lie algebra and $\manX=\Pi T
\manX_0$ with $\manX_0$ being a space-time manifold. If $e_i$ be a
basis in the Lie algebra then $c^i$ with $\p{c^i}=1$ are coordinates
on $\manM$. In addition we prescribe $\gh{c^i}=1$ and define
\begin{equation}
  Qc^i=\half\commut{e_j}{e_k}^i\,c^jc^k\,,
\end{equation}
which is the standard cohomology differential for a Lie algebra
$\algg$ with trivial coefficients. The variables $c^i$ are identified
with the ghost variables in the BRST formulation of the Lie algebra
cohomology. The BRST field theory described by $s$ is then a
non-Lagrangian version of the Chern-Simons theory. In particular, the
equations of motion are zero curvature equations for $\algg$-valued
connection 1-form. If in addition $\manX_0$ is a $3$-dimensional manifold and
$\algg$ is equipped with a nondegenerate invariant inner product then
the system is naturally Lagrangian and coincides with the AKSZ
formulation~\cite{\AKSZ} of the standard Chern-Simons theory.

\subsection{Linearization}

Suppose we are given with a map $\manX \to \manM$ defined by
$\psi^A=\psi_0^A(x,\theta)$ in terms of local coordinates. Let also
this map be such that $\derham \psi^A_0 +Q^A(\psi_0)=0$ i.e. it
determines a point on the zero locus of the differential $s$.  At
ghost number zero the configuration $\psi_0$ is a particular solution
to the equations of motion determined by $s$. Expanding the BRST
differential around a particular solution one gets
\begin{equation}
\label{eq:unf-s}
  s\psi^A
=\derham \psi^A+
\ddr{Q^A}{\psi^B}\Big|_{\psi=\psi_0}\,
\psi^B
+\half
\ffrac{\d_R^2 Q^A}{\d\psi^B\d\psi^C}\Big|_{\psi=\psi_0}
\psi^C\psi^B+\ldots\,,
\end{equation}
where $\ldots$ denote terms of higher orders in $\psi^A$.  In
particular, the linearized theory is determined by the following
linear differential
\begin{equation}
  s_0\psi^A=\derham \psi^A+\ddr{Q^A}{\psi^B}\Big|_{\psi=\psi_0}\,\psi^B\,.
\end{equation}

The linearized theory determined by $s_0$ can be identified with the
BRST field theory associated to the first-quantized system described
by the BRST operator
\begin{equation}
\label{eq:brst-op}
  \brst\phi=\derham\phi+ e_A \ddr{Q^A}{\psi^B}\Big|_{\psi=\psi_0}\, \phi^B\,,
\end{equation}
where $\phi=e_A \phi^A(x,\theta)$ is a general element of the ``space
of states'' which is the space of functions on $\manX$ with values in
the linear space $\cH$ identified with the tangent space to $\manM$.
More precisely, $e_A \phi^A(x,\theta)$ can be considered as a section
of the tangent bundle to $\manM$ pulled back by the map $\psi_0$.
From this point of view the BRST differential~\eqref{eq:unf-s} can be
naturally understood as that of a non-linear deformation of the linear
theory determined by the first-quantized BRST operator $\brst$.

The BRST operator~\eqref{eq:brst-op} has the same structure as that of
a parent systems constructed in~\cite{\BGST,\BGadS}. More generally,
one can consider a linear gauge field theory on $\manX_0$ whose BRST
differential have the form
\begin{equation}
\label{eq:s-form}
  s_0 \psi^A=\derham\psi^A+\bar\brst^A_B\psi^B\,,
\end{equation}
where $\bar\brst^A_B=\bar\brst^A_B(x,\theta)$ satisfies ``generalized zero
curvature'' condition 
\begin{equation}
\label{eq:brst-cond}
\derham\bar\brst^A_B+(-1)^{\p{A}+\p{C}}\bar\brst^A_C\bar\brst^C_B=0\,,
\end{equation}
needed for nilpotency.

The formulation where the BRST differential has the
form~\eqref{eq:s-form} with $\brst^A_B$
satisfying~\eqref{eq:brst-cond} can be considered as a BRST extension
of the linear unfolded
formulation~\cite{Vasiliev:1994gr,Vasiliev:1988xc}.  Indeed, if
$\bar\brst$ is a 1-form (i.e. is linear in $\theta^\mu$) and
$\gh{\psi^A}\geq 0$ then $\bar\brst$ can be considered a connection
1-form and equations of motion determined by $s_0$ take the form of a
covariant constancy condition
\begin{equation}
  d(\psi_p)^A+\bar\brst^A_B(\psi_p)^B=0\,.
\end{equation}
Here $\psi_p$ is a ghost-number-zero field entering
$\psi^A=(\psi_0)^A+\theta^\mu (\psi_1)_\mu^A+\theta^\mu
\theta^\nu(\psi_2)_{\mu\nu}^A+\ldots$, which is identified with a
$p$-form on $\manX_0$ with $p=\gh{\psi^A}$. From this perspective
parent theories constructed in~\cite{\BGST,\BGadS} are particular
examples of theories naturally emerging in the BRST extended unfolded
form.

\subsection{Consistent reductions}\label{sec:reduction}
Two local gauge field theories formulated within BRST framework are
naturally considered equivalent if they are related by
elimination/addition of generalized auxiliary fields.  Suppose that
after an invertible change of coordinates, possibly involving
derivatives, the set of fields $\psi^A$ splits into
$\varphi^\alpha,w^a,v^a$ such that equations $sw^a|_{w^a=0}=0$
(understood as algebraic equations in the space of fields and their
derivatives) are equivalent to $v^a=V^a[\varphi^\alpha]$, i.e., can be
algebraically solved for fields $v^a$. Fields $w,v$ are then
generalized auxiliary fields. The field theory described by $s$ is
equivalent to that described by the reduced differential $\tilde s$
acting on the space of fields $\varphi^\alpha$ and their derivatives
and defined by $\tilde s \varphi^\alpha=s\varphi^\alpha|_{w^a=0,\,
  v^a=V^a[\varphi]}$ (see \cite{\BGST} for more details). In the
Lagrangian framework, fields $w,v$ are in addition required to be
second-class constraints in the antibracket sense.  In this context,
generalized auxiliary fields were originally proposed
in~\cite{Dresse:1990dj}. Generalized auxiliary fields comprise both
standard auxiliary fields and pure gauge degrees of freedom as well as
their associated ghosts and antifields.

For BRST field theory $(\manX,\derham,\manM,Q)$ one easily finds
generalized auxiliary fields as originating from contractible pairs
for $Q$. Namely, let $w^a$ be such that $w^a,Qw^a$ are independent
constraints on $\manM$ determining the submanifold $\tilde\manM\subset
\manM$. The theory $(\manX,\derham,\manM,Q)$ is then equivalent via
elimination of generalized auxiliary fields to
$(\manX,\derham,\tilde\manM,\tilde Q)$ with $\tilde Q=Q|_{\tilde\manM}$. In order
to see that $Q$ indeed restricts to $\tilde\manM$ it is enough to observe
that $(Qw^a)|_{\tilde\manM}=0$ and $Q(Qw^a)=0$. For more details we refer
to~\cite{\BGunf}.

Analogously, one can consider contractible pairs $t^\alpha$ and
$\derham t^\alpha$ in the extended space-time manifold $\manX$.
Namely, suppose that $t^\alpha$ and $\derham t^\alpha$ are independent
regular constraints determining a submanifold $\tilde\manX$. One can
address the question on the relation of $(\manX,\derham,\manM,Q)$ and
$(\tilde\manX,\tilde\derham,\manM,Q)$ where
$\tilde\derham=\derham|_{\tilde\manX}$.  These theories can not be
considered equivalent as local field theories because they live on
different space-time manifolds. However, if the coordinates
transversal to $\tilde\manX\subset \manX$ are considered as internal
degrees of freedom rather than space-time coordinates one can indeed
show that respective theories are equivalent.  For more details we
again refer to~\cite{\BGunf}. In particular, if $\manX=\Pi T^*\manX_0$
with coordinates $x^\mu,\theta^\mu$ one can consistently eliminate any
pair $x^\nu,\theta^\nu$.  Note that the auxiliary role of space-time
coordinates was observed in~\cite{Vasiliev:2001zy,Gelfond:2003vh} in
the context of HS theories formulated within unfolded framework.

If one is given with a particular solution $\Psi_0$ satisfying
$\derham\psi^A_0+Q^A(\psi_0)=0$ then the system expanded around
$\Psi_0$ can be reduced using the reduction machinery developed
in~\cite{\BGST,\BGadS} for the free theories associated to
fist-quantized systems. Indeed, the linearized theory can be
identified with the free field theory associated to the
first-quantized system described by~\eqref{eq:brst-op}.  Under the
standard assumptions it then follows that the generalized auxiliary
fields for the linearized theory are also generalized auxiliary fields
for its nonlinear deformation (see e.g.~\cite{\BGadS}).

\subsection{Putting a quantum constrained system to a fiber}
\label{sec:general}

Let us consider again a constrained system quantized in the sense of
deformation quantization, i.e. the associative $*$-product algebra
$\algA$ of the extended phase space functions depending formally on
$\hbar$ and equipped with the ghost number grading and the Grassmann
parity. Contrary to the construction of Section~\bref{sec:basic-1} now
we are going to achieve a generally covariant (in the sense of
$\manX_0$) description of the theory.  To this end we construct an
AKSZ-type sigma-model by, roughly speaking, putting the quantum
constrained system to the target space. Moreover, we need to change
the class of the phase space functions. Namely, we replace the
space-time coordinates $x^\mu$ with the formal variables $y^a$ so that
$\algA$ consist of formal power series in $y^a$ and $\hbar$ with
coefficients depending on the remaining variables. In addition, we
also assume $\algA$ to be graded-finite dimensional.

More technically, we first consider a supermanifold associated to
$\algA$.  Let $e_A$ be a basis in $\algA$ and $\psi^A$ coordinates on
the associated supermanifold $\manM$. The string field is then given
by $\Psi=e_A \psi^A$. Similarly to the considerations
in~\bref{sec:basic-1} the $Q$-structure on $\manM$ is given by
\begin{equation}
Q\psi^A=\bar\brst^A_B\psi^B+\ffrac{1}{2\hbar}  \qcommut{e_B}{e_C}^A \,\psi^C \psi^B(-1)^{\p{B}}\,.
\end{equation}
where $\qcommut{f}{g}=f*g-(-1)^{\p{f}\p{g}} g* f$ is a $*$-commutator
in $\algA$ and $\bar\brst\map\algA\to\algA$ a ``fiber'' BRST operator
$\bar\brst e_A=e_B\bar\brst^B_A$ satisfying
$\bar\brst^2=0,\,\gh{\bar\brst}=1,\,\p{\bar\brst}=1$ and
$\bar\brst(\phi*\chi)=
(\bar\brst\phi)*\chi+(-1)^{\p{\phi}}\phi*(\bar\brst\chi)$.  In terms
of the string field the definition of $Q$ takes the form
\begin{equation}
Q\Psi=\bar\brst\Psi+\ffrac{1}{2\hbar}\qcommut{\Psi}{\Psi}\,.
\end{equation}
Note that the construction can be naturally generalized to involve
$L_\infty$-structure instead of a differential graded Lie algebra.
This corresponds to taking odd nilpotent vector field $Q$ not
necessarily quadratic in $\psi^A$ (see~\cite{\BGunf} for more
details).

Given a supermanifold $\manX$ equipped with the differential $\derham$
one can then build a BRST system $(\manX,\derham,\manM,Q)$ whose BRST
differential is determined by 
\begin{equation}
\label{eq:s-def2}
s\Psi=\derham \Psi+\bar\brst\Psi+\ffrac{1}{2\hbar}\qcommut{\Psi}{\Psi}\,,
\end{equation}
For definiteness, we take $\manX=\Pi T\manX_0$ and $\derham$ to be de
Rham differential $\theta^\mu \dl{x^\mu}$ with $x,\theta$ being
coordinates on $\manX_0$ and fibers of $\Pi T\manX_0$. But the general
considerations remain the same in the case where $\manX$ is a general
supermanifold equipped with the ghost degree, the odd nilpotent
vector field $\derham,\,{\mathrm gh}(\derham)=1$, and the volume form.
As was discussed above this also concerns the generalization to the
case where $\algA$ is identified with the fiber of a nontrivial vector
bundle over $\manX$ with the transition function preserving the
$Q$-structure.

Similarly to the Chern-Simons theory example, the $Q$ structure on
$\manM$ is nothing but the standard cohomology differential for
$\algA$ considered as a differential graded Lie algebra.  Variables
$\psi^A$ can also be identified with the respective ``ghost''
variables. However, contrary to the case of Chern-Simons theory
$\psi^A$ can have arbitrary ghost degree depending on $\gh{e_A}$. In
particular, it follows from counting the ghost degree that the
physical fields entering $\psi^A(x,\theta)$ appear to be differential
forms of form degree depending on $\gh{\psi^A}$ and not necessarily
$1$-forms as in the case of Chern-Simons theory.

Suppose we are interested in a particular $\Psi_0=e_A\psi^A_0(x,\theta)$
satisfying $(s\Psi)|_{\Psi=\Psi_0}=0$.  Among possible particular
solutions there is a class of solutions which are in some sense
natural. These can be identified if one observes that equation
$(s\Psi)|_{\Psi=\Psi_0}=0$ for this system can be considered as the
quantum master equation (in the $*$-product sense) for the
Fedosov-like extension of the constrained system $\algA$.  Indeed, let us introduce momenta
$\bar p_\mu,\cP_\mu$ conjugates to $x^\mu,\theta^\mu$ so that at the
quantum level
\begin{equation}
\qcommut{x^\mu}{\bar p_\nu}=\hbar \delta^\mu_\nu\,,\qquad
\qcommut{\theta^\mu}{\cP_\mu}=-\hbar\delta^\mu_\nu\,,
\end{equation}
and extend $\algA$ by the $*$-product algebra generated by
$x^\mu,\theta_\mu,\bar p_\mu,\cP_\mu$ (in the case where $\manX_0$ is
a curved manifold one needs to consider the star product algebra
arising in quantization of $T^*(\Pi T\manX_0)$).  Identifying $\Psi_0$
as a ghost-number-one phase space function one observes that the
quantum master equation
$\bar\brst\Psi^\prime_0+(2\hbar)^{-1}\qcommut{\Psi^\prime_0}{\Psi^\prime_0}=0$
for a quantum BRST charge $\Psi^\prime_0=\Psi_0-\theta^\mu \bar p_\mu$
is indeed equivalent to $(s\Psi)_{\Psi=\Psi_0}$. Note that if
$\bar\brst=\hbar^{-1}\qcommut{\bar\Psi_0}{\cdot}$ for some
$\bar\Psi_0\in \algA$ then the master equation for
$\Psi^{\prime\prime}_0=\Psi_0+\bar\Psi_0-\theta^\mu \bar p_\mu$ takes
the standard form $\qcommut{\Psi^{\prime\prime}_0}{\Psi^{\prime\prime}_0}=0$.  Under the
appropriate regularity assumptions one can also show that all physical
quantities (representatives of the BRST cohomology) can be assumed
independent on $\bar p_\mu$ and $\cP_\mu$ so that it is legitimate to
eliminate them from the formulation.

Given a particular solution $\Psi_0$ the BRST differential of
the theory expanded around $\Psi_0$ is determined by
\begin{equation}
  s\Psi=(s_0+s^\prime)\Psi=\derham\Psi+\bar\brst\Psi
+\ffrac{1}{\hbar}\qcommut{\Psi_0}{\Psi}+\ffrac{1}{2\hbar}\qcommut{\Psi}{\Psi}\,,
\end{equation}
where
$s_0\Psi=\derham\Psi+\bar\brst\Psi+\hbar^{-1}\qcommut{\Psi_0}{\Psi}$
is the BRST differential of the linearized theory.  As in the general
case considered above the BRST field theory determined by $s_0$ can be
identified with that associated with the first-quantized system
$(\brst,\Gamma(\bundle\algA,\manX))$ on $\manX$ with the space of
states being $\Gamma(\bundle\algA,\manX)$ (i.e. sections of
$\bundle\algA$ considered as a bundle over $\manX$) and the BRST
operator $\brst$ determined by
$\brst\phi=\derham\phi+\bar\brst\phi+\hbar^{-1}\qcommut{\Psi_0}{\phi}$.
It can be more natural to interpret this system as
$(\brst,\Gamma(\bundle\algA\tensor \Lambda,\manX_0))$ defined on
$\manX_0$.  In this case the space of states becomes
$\Gamma(\bundle\algA\tensor\Lambda,\manX_0)$, where $\Lambda$ denotes
the Grassmann algebra generated by $\theta^\mu$.  Indeed, elements of
$\Gamma(\bundle\algA\tensor\Lambda,\manX_0)$ over $\manX_0$ are
identified with elements of $\Gamma(\bundle\algA,\manX)$ over $\manX$,
which, in turn, are naturally considered as differential forms on
$\manX_0$ with values in $\algA$.

Note, that similarly to Section~\bref{sec:basic} all the present
considerations also remain true if one takes the classical limit
$\hbar\to 0$ in the expressions for $s_0$ and $\brst$ and considers
$\algA$ as a commutative algebra equipped with the Poisson bracket.

It is important to stress that a priori one can take arbitrary
$Q$-manifolds $\manM$ and $\manX$ in order to construct some BRST
field theory $(\manX,\derham,\manM,Q)$. However, this theory can
usually be interpreted in terms of some meaningful model if one
specifies (a class of) $\bar\brst$ operators and/or particular
solutions $\Psi_0$ involving e.g.  some geometrical structures on
$\manX_0$. In particular, this can fix the geometry and dimension of
$\manX_0$ (see e.g.~\cite{Vasiliev:2001zy,Gelfond:2003vh} where a
similar ideas have been utilized in the context of HS theory). The
advantage of the BRST theory approach developed here is that it
provides a guiding rule by interpreting the particular solutions for
the system $(\manX,\derham,\manM,Q)$ as the solution to the master
equation for some (quantum) constrained system.

The following example provides a simple illustration and at the same
time demonstrates the flexibility of the construction: let $\manX_0$ be
a symplectic manifold and $\algA$ be the algebra of formal power
series in $y^a$ identified with the coordinates on the symplectic
vector space isomorphic to a tangent space $T_p\manX_0$ at
$p\in\manX_0$. Equation
$\derham\Psi_0+(2\hbar)^{-1}\qcommut{\Psi_0}{\Psi_0}=0$ then coincides
with the zero curvature condition in the Fedosov
quantization~\cite{Fedosov:1994} provided one imposes suitable
additional conditions on $\Psi_0$. This is not surprising because as
it was shown in~\cite{Grigoriev:2000rn} the Fedosov quantization
itself can be understood as a BRST quantization of some specially
prepared constrained system. The case where $\algA$ also contains
ghost variables corresponds to the extension of the Fedosov
quantization to the case of systems with constraints.

\section{Off-shell higher spin fields}
\label{sec:HS}
As an illustration let us consider the simplest constrained system
from Section~\bref{sec:basic-example} but contrary
to~\bref{sec:basic-example} we do not embed the space-time manifold
$\manX_0$ into the extended phase space.  Namely, we take the phase
space to be $T^* V$, where $V$ is a linear $\dmn$-dimensional space.
The coordinates on $T^*V$ are $y^a,p_a,\,$ $a=1,\ldots, \dmn$, with
$y^a$ being coordinates on $V$ and $p_a$ its conjugate momenta. The
Weyl star product on the phase space is determined by
$\qcommut{y^a}{p_b}=\hbar\delta^a_b$.

As before the phase space and the star product is further extended by
the appropriate Grassmann odd ghost variables $c,\pi$ with
$\qcommut{\pi}{c}=-\hbar$ and $\gh{c}=1,\,\gh{\pi}=-1$. According to the general construction
described in~\bref{sec:general} we now take $\algA$ to be the algebra
of formal power series in $y^a$ with coefficients being polynomials in
$p_a,c,\pi$. One then associates supermanifold $\manM$ with $\algA$
and introduces the component fields so that the string field $\Psi$
takes the form
\begin{equation}
\Psi= \bA(x,\theta;y,p)+c \bF(x,\theta;y,p)+\pi\bR(x,\theta;y,p)+ c \pi \bB(x,\theta;y,p)\,.
\end{equation}
It follows from $\gh{\Psi}=\p{\Psi}=1$ that
\begin{equation}
  \gh{\bA}=\gh{\bB}=1\,, \qquad \gh{\bR}=2\,,\qquad \gh{\bF}=0
\end{equation}
while the Grassmann parity is just a ghost number modulo $2$. 

As a supermanifold $\manX$ we take $\Pi T\manX_0$, with $\manX_0$ being
the manifold with the tangent space isomorphic $V$. Throughout this
Section we do not assume that $T\manX_0$ is a trivial bundle (i.e.
$\manX_0$ is parallelizable) because the required generalization of the
construction is straightforward. Indeed, in this case the bundle with
the fiber $\algA$ is obviously just a vector bundle associated with
$T\manX_0$ (sections of $\bundle\algA$ are identified with tensor
fields on $\manX_0$).

Consider then the BRST field theory $(\manX,\derham,\manM,Q)$ determined
by the data above. The BRST differential encoding the equations of
motion and gauge symmetries is determined by
$s\Psi=\derham\Psi+Q\Psi=\derham\Psi+(2\hbar)^{-1}\qcommut{\Psi}{\Psi}$.
Note that in this example we took $\bar\brst=0$. However, expanding
the theory around a particular solution $\Psi_0$ one can always
introduce $\bar\brst=\qcommut{\bar\Psi_0}{\cdot}$ by replacing
$\Psi_0$ with $\Psi_0+\bar\Psi_0$.

In order to demonstrate explicitly the structure of equations of
motion and gauge symmetries let us spell them out in terms of the
component fields. It is useful to represent the string field as
$\Psi=\Psi(x,\theta^\mu;y,p,c,\pi)$. Physical fields are identified
with ghost-number-zero component fields entering $\Psi$ and are given
by zero form $F(x;y,p)$, 1-form $A_\mu(x;y,p)$, 2-for
$R_{\mu\nu}(x;y,p)$, and 1-form $B_\mu(x;y,p)$. These are particular
component fields entering $\bF,\bA,\bR$ and $\bB$ respectively. In
terms of component fields the equations of motion determined by the
BRST differential read as
\begin{equation}
\begin{aligned}
\label{eq:ext-basic}
\derham A+\ffrac{1}{2\hbar}\qcommut{A}{A}+\ffrac{\hbar}{8}\qcommut{B}{B}-\ffrac{1}{2}\scommut{F}{R}&=0\,,\\
\derham B+\ffrac{1}{\hbar}\qcommut{A}{B}+\ffrac{1}{2\hbar}\qcommut{F}{R}&=0\,,\\
\derham F+\ffrac{1}{\hbar}\qcommut{A}{F}-\ffrac{1}{2}\scommut{F}{B}&=0\,,\\
\derham R+\ffrac{1}{\hbar}\qcommut{A}{R}+\ffrac{1}{2}\scommut{R}{B}&=0\,,
\end{aligned}
\end{equation}
where $\scommut{a}{b}=a*b+(-1)^{\p{a}\p{b}} b*a$ denotes the graded
anticommutator.

The gauge symmetries determined by the BRST differential have the form
$\delta\Psi^{(0)}=\derham\Lambda+\hbar^{-1}\qcommut{\Psi^{0}}{\Lambda}$, where
$\Lambda$ is $\Psi^{(1)}$ with the component fields (which are of
ghost number one) replaced with the gauge parameters depending arbitrary
on $x$. The gauge transformations take the form
\begin{equation}
\label{eq:GS-ext}
  \begin{aligned}
    \delta_\Lambda A&=\derham \lambda+\ffrac{1}{\hbar}\qcommut{A}{\lambda}+\ffrac{1}{2}\scommut{F}{\xi}
             +\ffrac{\hbar}{4}\qcommut{B}{\chi}\,,\\
    \delta_\Lambda B&=\derham \chi+\ffrac{1}{\hbar}\qcommut{A}{\chi}+\ffrac{1}{\hbar}\qcommut{B}{\lambda}
             -\ffrac{1}{2}\scommut{F}{\xi}\,,\\
    \delta_\Lambda R&=\derham \xi+\ffrac{1}{\hbar}\qcommut{A}{\xi}+\ffrac{1}{\hbar}\qcommut{R}{\lambda}
             -\ffrac{1}{2}\scommut{R}{\chi}+\ffrac{1}{2}\scommut{B}{\xi}\,,\\
    \delta_\Lambda F&=\ffrac{1}{\hbar}\qcommut{F}{\lambda}+\ffrac{1}{2}\scommut{F}{\chi}\,,
\end{aligned}
\end{equation}
where $\Lambda=\lambda-b_0\xi+c_0b_0\chi$ (note that the term
proportional to just $c_0$ is missing because the respective term in
the string field does not enter $\Psi^{(1)}$ as it follows from
counting ghost degree). It also follows from counting ghost degree
that $\lambda$, $\xi$ and $\chi$ are respectively $0$, $1$, and
$0$-forms.

If one puts $R=B=0$ and restricts the gauge parameter such that $\xi=\chi=0$ the system described
by~\eqref{eq:ext-basic} and \eqref{eq:GS-ext} reduces to that  recently considered
by M.~Vasiliev in~\cite{Vasiliev:2005zu}. The equations of motion and gauge symmetries take the form
\begin{align}
\label{eq:basic-gen-A}
  \derham A+\frac{1}{2\hbar}\qcommut{A}{A}=0\,,\\
\label{eq:basic-gen-F}
 \derham F+\frac{1}{\hbar}\qcommut{A}{F}=0\,,
\end{align}
and
\begin{equation}
    \delta_\lambda A=\derham \lambda+\ffrac{1}{\hbar}\qcommut{A}{\lambda}\,,\qquad 
    \delta_\lambda F=\ffrac{1}{\hbar}\qcommut{F}{\lambda}\,,
\end{equation}
respectively.

It was shown in~\cite{Vasiliev:2005zu} that when expanded around the
particular solution corresponding to the Minkowski space-time this
system gives a non-linear off-shell description of the HS gauge fields
without the trace constraint. As we are going to see, the analogous
expansion of the BRST field theory $(\manX,\derham,\manM,Q)$ properly
describes off-shell HS fields with the trace constraint taken into
account.  In this description additional fields and gauge symmetries
entering component equations \eqref{eq:ext-basic} and
\eqref{eq:GS-ext} effectively eliminate the traces at the non-linear
level.

Equations~\eqref{eq:basic-gen-A}-\eqref{eq:basic-gen-F} can also be
understood as the basic equations of the particular version of the
Fedosov quantization~\cite{Fedosov:1994}.  More precisely, the
version~\cite{Bordemann:1997er} adapted to the case where the phase
space is a cotangent bundle. In particular, \eqref{eq:basic-gen-A} is
nothing but the vanishing curvature condition for a nonlinear
connection while~\eqref{eq:basic-gen-F} is a covariant constancy
condition for observable $F$.\footnote{The author wants to thank
  A.~Sharapov for sending his unpublished work with A.~Segal where the
  analogous equations have been also considered in the context of
  conformal HS theory.}  As we are going to see, a more general point
of view is to consider the
system~\eqref{eq:basic-gen-A}-\eqref{eq:basic-gen-F} as a truncated
version of the bigger system~\eqref{eq:ext-basic}. In its turn this
system can be understood as a component form of the BFV-BRST quantum
master equation for the particular quantum constrained system.

\subsection{Linearization around Minkowski space}\label{sec:flat}
As a simplest case to begin with we take $\manX_0$ to be flat
Minkowski space and study the linearization of the system
$(\manX,\derham,\manM,Q)$ around a particular solution that describes
a scalar particle on $\manX_0$. A useful choice of the particular
solution to $\derham\Psi_0+(2\hbar)^{-1}\qcommut{\Psi_0}{\Psi_0}=0$ is
then $\Psi_0=A_0+cF_0$, where
\begin{equation}
  A_0=\theta^\mu e_\mu^a p_a+\omega_{\mu b}^a y^b p_a\,, \qquad 
F_0=\hhalf\eta^{ab}p_a p_b\,, \qquad R_0=0\,,\qquad B_0=0\,,
\end{equation}
all the fields in nonzero ghost degree vanish, and $e,\omega,\eta$
(identified with the coefficients of the flat vielbein,the Lorentz
connection, and the flat Minkowski metric on $V$) satisfy
\begin{equation}
 de^a+\omega^a_b e^b=0\,, \qquad
  d\omega^a_b+\omega^a_c\omega^c_b=0\,,\qquad 
  \omega^a_c\eta^{cb}+\eta^{ac}\omega_c^b=0\,.
\end{equation}
Indeed, $\Psi_0$ coincides with the quantum BRST charge of the parent
system~\cite{\BGST} for the scalar particle on the Minkowski space (of
course, one also needs to add $-\theta^\mu\bar p_\mu$ to take
$\derham$ into account, see Section~\bref{sec:general}).  Although
this observation looks too obvious in the case at hand it can be
helpful in identifying particular solutions in more involved
situations.

The BRST differential of the linearized theory is given by
$s_0\Psi=\derham \Psi+\hbar^{-1}\qcommut{\Psi_0}{\Psi}$.  Using the
explicit expression for $\Psi_0$ one finds
\begin{equation}
\label{eq:Mink-brst}
  s_0\Psi=\brst\Psi\,,\qquad 
\brst=\nabla + \sigma-c p^a\dl{y^a} -\half p^2\dl{\pi}
    -\ffrac{\hbar^2}{8}\ffrac{\d^2}{\d y^a\d y_a}\dl{\pi}\,,
\end{equation}
where we have introduced the following notations:
\begin{equation}
\label{eq:nabla}
  \nabla=\derham-\theta^\mu \omega_{\mu b}^a y^b\dl{y^a}+
\omega_{\mu b}^a p_a\dl{p_b}\,, \qquad \sigma=-\theta^\mu e_\mu^a\dl{y^a}\,.
\end{equation}
We consider the linearized theory described by $s_0$ as the field
theory associated to the first-quantized system
$(\brst,\Gamma(\bundle{\algA},\manX))$.  Recall that in the case at hand
$\algA$ is the algebra of formal power series in $\hbar$ with
coefficients in functions in $y,p,c,\pi$ which are formal power series
in $y$ and polynomials in the rest of variables.

To see that $s_0$ indeed describes off-shell version of Fronsdal HS
theory let us reduce the system $(\brst,\Gamma(\bundle\algA,\manX))$ to that
with the traceless fields.  Namely, we take as a degree homogeneity in
$\pi$ so that $\brst=\brst_{-1}+\brst_{0}$ with $\brst_{-1}=-(\hhalf
p^2+\ffrac{\hbar^2}{8}\Box)\dl{\pi}$ (see \cite{\BGST,\BGadS} for
details on the consistent reductions in homological terms). Cohomology
of $\brst_{-1}$ in $\algA$ is concentrated in degree zero and can be
identified with the subspace $\cE\subset \algA$ of $\pi$-independent
elements annihilated by $T=\dl{p^a}\dl{p_a}$ (i.e.  traceless
elements). This is the standard fact in zeroth degree in $\hbar$.  In
order to show this to all orders in $\hbar$ one needs to subtract
traces order by order in $\hbar$.

Because the cohomology is concentrated in one degree only the reduction is
straightforward and one immediately arrives at the reduced system
$(\tilde\brst,\Gamma(\bundle\cE,\manX))$ with $\tilde\brst$ being
$\brst_0$ projected to $\bundle\cE$
\begin{equation}
  \tilde\brst=\nabla+\sigma-c\, \Pj^\prime_{\T} p^a\dl{p^a}\,.
\end{equation}
Here $\Pj^\prime_{\T}$ denotes the projector to $\cE$ defined as
follows: for $\phi=\phi_0+(p^2+\ffrac{\hbar^2}{8}\Box)\phi^\prime$
with $T\phi_0=0$ one defines $\Pj^\prime_{T}\phi=\phi_0$. Note that in
the classical limit $\hbar \to 0$ this $\Pj^\prime_{\T}$ reduces to
the standard projector $\Pj_{\T}$ to the traceless component.

If one takes $\Psi^{(0)}=\theta^\mu A_\mu+c F$ the equations of motion
$\tilde\brst\Psi^{(0)}=0$ of the associated field theory take the form
\begin{equation}
\label{eq:off-flat}
    (\nabla +\sigma) A=0\,,\qquad (\nabla +\sigma)  F=- \Pj^\prime_{\T} p^a\dl{y^a}  A\,.
\end{equation}
while the gauge symmetry $\delta_\lambda\Psi^{(0)}=\tilde\brst\Psi^{(1)}$
(with ghost-number-one fields in $\Psi^{(1)}$ replaced with the gauge parameters)
read as
\begin{equation}
\label{eq:off-flat-gs}
  \delta A=(\nabla+\sigma)\lambda\,,\qquad \delta F=-\Pj^\prime_{\T} p^a\dl{y^a}\lambda\,,
\end{equation}
where $\lambda=\lambda(x;y,a)$ is a gauge parameter satisfying
$T\lambda=0$.

The system can be consistently restricted to
$(\brst_{\os},\Gamma(\bundle\cE_{\os},\manX))$, with
\begin{equation}
  \brst_\os=\nabla+\sigma-c p^a\dl{y^a}
\end{equation}
and $\cE_{\os}\subset \cE$ being the subspace of elements annihilated
by
\begin{equation}
S=\ffrac{\d^2}{\d p_a \d y^a}\,, \qquad \Box=\ffrac{\d^2}{\d y_a \d y^a}\,.
\end{equation}
Note that projector is not anymore needed because $p^a\dl{y^a}\phi$
belongs to $\cE_\os$ provided $\phi\in \cE_\os$.

Contrary to the equivalent reduction considered above this restriction
imposes dynamical equations. Namely, the system
$(\brst_{\os},\Gamma(\bundle\cE_{\os},\manX))$ explicitly coincides with
so-called intermediate system originally constructed in~\cite{\BGST},
where it was shown to properly describe Fronsdal HS theory on the
Minkowski space. One then concludes that the theory determined by
$s\Psi=\brst\Psi+(2\hbar)^{-1}\qcommut{\Psi}{\Psi}$ indeed describes
the non-linear deformation of the linear off-shell HS theory on the
Minkowski space.  At the level of the associated field theory this can
be easily observed as follows: if $A,F,\lambda$ in~\eqref{eq:off-flat}
and \eqref{eq:off-flat-gs} in addition satisfy $\Box A=SA=0$, $\Box F=S
F=0$, and $\Box \lambda=S \lambda =0$ then \eqref{eq:off-flat} and
\eqref{eq:off-flat-gs} are just equations of motion and gauge
symmetries of the Fronsdal theory in the intermediate
form~\cite{\BGST}.

The off-shell theory described by $s_0$ is equivalent to the off-shell
theory constructed in Section~\bref{sec:basic-example} through the
elimination of generalized auxiliary fields.  Indeed, it was shown
in~\cite{\BGST} that if the BRST operator has the
form~\eqref{eq:Mink-brst} then it describes the parent system
constructed for the first-quantized BRST system with 
\begin{equation}
\brst_{non-extended}=
-c p^\mu\dl{x^\mu}-(p^2+\ffrac{\hbar^2}{8}\dl{x^\mu}\dl{x_\mu})\dl{\pi}\,
\end{equation}
(here, $x^\mu$ are standard flat coordinates on Minkowski space) which
is the BRST operator describing the linearized off-shell theory
considered in~\bref{sec:basic-example}. More precisely, the BRST
operator encodes the linearizzation of gauge
tranformations~\eqref{eq:gs-particle}.

Several comments are in order: one can equally consider the classical
version of the system $(\manX,\derham,\manM,Q)$.  This gives in
general different linearized off-shell system which can be obtained by
taking $\hbar\to 0$ limit.  In fact both linearized systems coincide
on-shell because the quantum corrections to $\brst$ involve operator
$\Box$ that vanishes on-shell.

Another way to see that $(\brst,\Gamma(\bundle\algA,\manX))$ describes
off-shell HS fields in Minkowski space is to incorporate the
constraints $\Box,S$ enforcing dynamical equations in the BRST
operator. More precisely, one needs to introduce new ghost pairs
$c_0,b_0$ and $\cd,b$ (extending therefore $\algA$ to $\algA^\T$) and
to consider extended BRST operator
\begin{equation}
\label{eq:brst-t}
\brst^{\T}=\brst+c_0\Box-\cd S+\text{terms cubic in ghosts}\,.
\end{equation}
One then finds that $\brst^{\T}$ is the BRST operator of the parent
system~\cite{\BGST} provided one redefines constraints in order to get
rid of the term proportional to $\hbar^2\Box$ and identify $p$ with
$a$ and $\dl{p}$ with $-a^\dagger$. The only difference is that
in~\cite{\BGST} oscillators $a,a^\dagger$ are represented on
polynomials in $a^\dagger$ and the representation for some ghost pairs
is also chosen differently.~\footnote{The choice of representation for
  Grassmann odd ghost variables is not really important because all
  ``good'' ones are equivalent.}  Nevertheless, one can directly check
that the system $(\brst^\T,\Gamma(\bundle\algA^\T,\manX))$ is just a
different realization of the same parent system. In fact, the relation
between these realizations can be understood in terms of the
appropriate automorphism of the $sp(4)$ algebra (see~\cite{\BGST})
inducing the exchange $a \to \dl{a}$, $\dl{a} \to -a$ in the
representation.  We then conclude that the system
$(\brst,\Gamma(\bundle\algA,\manX))$ is a truncated version of the
parent system for Fronsdal HS fields.  In particular, $\brst$ can be
directly obtained from the parent BRST operator by redefining
constraints and dropping the terms with constraints $\Box,S$ and the
respective ghost variables.

The off-shell theory determined by $s_0$ can be also understood as an
off-shell formulation not only for the Fronsdal model. Indeed, since it is
related through the elimination of generalized auxiliary to the
off-shell theory from Section~\bref{sec:basic-example} it can be also
used to describe conformal HS fields.

\subsection{Linearization around arbitrary background}

Let us now turn to the case where $\manX_0$ is not necessarily a
Minkowski space and assume for the moment that $\manX_0$ is a general
(pseudo)Riemannian manifold. As before local coordinates on $\manX_0$ are
denoted by $x^\mu$. In this more general setting one can also address
the question on particular solutions to~\eqref{eq:ext-basic}.  It is
again enough to restrict to the class of solutions with $R=B=0$ so
that the remaining equations are~\eqref{eq:basic-gen-A} and
\eqref{eq:basic-gen-F}.

In order to analyze equations \eqref{eq:basic-gen-A} and
\eqref{eq:basic-gen-F} for general $\manX_0$ it is useful to prescribe
the following degrees to the variables
\begin{align}
  \deg{x}=\deg{\theta}=\deg{p}=0\,,\qquad \deg{y}=1\,,\qquad
  \deg{\hbar=1}\,,
\end{align}
so that the $*$-product carries vanishing degree.  Let $A=\sum_n
A_{[n]}$ with $\deg{A_{[n]}}=n$. Suppose that the following boundary
conditions are imposed on $A$
\begin{equation}
\label{eq:bc1}
  A_{[0]}=\theta^\mu e_\mu^a p_a\,,\qquad  A_{[1]}=\theta^\mu \omega_{\mu a}^b y^a p_b\,,
\end{equation}
where $e_\mu^a$ is assumed nondegenerate and $e,\omega$ are
compatible, i.e. $de^a+\omega^a_b e^b=0$. The geometric interpretation
of $e,\omega$ is standard: $e$ is a vielbein (i.e. it determines an isomorphism between the
tangent bundle and the bundle with the fiber $V$) and $\omega$ is a
connection $1$-form. One has the following
\begin{prop}\cite{Fedosov:1994,Bordemann:1997er}
  Given a nondegenerate vielbein $e$ and a connection $\omega$ there
  exists $A=\theta^\mu A_\mu(x;y,p)$ satisfying $\derham A+
  (2\hbar)^{-1}\qcommut{A}{A}=0$ and boundary
  conditions~\eqref{eq:bc1}. Under the additional condition $y^a
  e_a^\mu \dl{\theta^\mu} A_{[n]}=0$ for $n \geq 2$ the solution is
  unique. Any other solution satisfying the same boundary condition
  can be obtained by a gauge transformation.
\end{prop}
\begin{rem}
  For any $e,\omega$ the solution can always be taken linear in $p$.
  In particular such a solution also satisfies the classical version
  of equation~\eqref{eq:basic-gen-A}, i.e., with the $*$-commutator
  replaced with the Poisson bracket.
\end{rem}
A geometrical meaning of this statement is that any background
geometry described by a vielbein and a (not necessary flat) torsion-free linear
connection can be described by a flat non-linear connection.

Suppose that we have a particular solution $A$ satisfying
\eqref{eq:basic-gen-A} and \eqref{eq:bc1}. Let us now turn to the
equation~\eqref{eq:basic-gen-F}. 
\begin{prop}\cite{Fedosov:1994,Bordemann:1997er}
  Let $F_{[0]}(x;p)$ be an arbitrary polynomial in $p$ with coefficients
  being tensor fields on $\manX_0$.  There exists unique $F(x;y,p)$
  satisfying $\derham F+\hbar^{-1}\qcommut{A}{F}=0$ and the boundary
  condition
  \begin{equation}
    F(x;0,p)=F_{[0]}(x;p)\,.
  \end{equation}
\end{prop}
\begin{rem}
  If $A$ is linear in $p_a$ then the Proposition also applies to the
  Poisson bracket counterpart of equation~\eqref{eq:basic-gen-F}.
  Moreover, in this case if $F_{[0]}$ is homogeneous in $p$ then $F$
  satisfying the Poisson bracket version of the Proposition is also
  homogeneous in $p$ of the same order.
\end{rem}

Suppose $A_0,F_0$ be a particular solution
to~\eqref{eq:basic-gen-A}-\eqref{eq:basic-gen-F} subjected to the
boundary condition
\begin{equation}
  A_0=e^ap_a+\omega^a_b y^bp_a+\ldots\,, \qquad F_0=\hhalf\eta^{ab}p_a p_b+\ldots\,,
\end{equation}
where $\ldots$ denote higher degree terms and $\omega,\eta$ in
addition satisfy $\omega^c_a\eta_{cb}+\eta_{ac}\omega^c_b=0$. By
expanding the theory around $\Psi_0=A_0+c F_0$ one arrives at the
theory described by the BRST differential $s\Psi=\derham
\Psi+{\hbar}^{-1}\qcommut{\Psi_0}{\Psi}+(2\hbar)^{-1}\qcommut{\Psi}{\Psi}$.
This theory can be considered as a natural generalization of the
off-shell HS theory constructed in Section~\bref{sec:flat} to the case
where the background manifold is an arbitrary (pseudo)Riemannian
manifold. However, it is important to note that even a linearization
of this theory has, in general, nothing to do with the conventional
Fronsdal HS theory because Fronsdal HS fields can live only on
constant curvature spaces, i.e.  there should be obstructions in
putting the theory on-shell.

In the case where $\manX_0$ is the constant curvature space one can
easily construct a particular solution and consider the respective
linearized theory. However, we do no take this way in the present
paper. Instead, in Section~\bref{sec:ads} we describe the off-shell HS
theories on constant curvature backgrounds in terms of the embedding
space.

\section{Off-shell HS fields on AdS in terms of the embedding space}
\label{sec:ads}

\subsection{Non-linear off-shell HS fields on AdS}
\label{sec:HS-ads}
Following~\cite{\BGadS} let $Y^A$ be coordinates on
$\dmn+1$-dimensional pseudo-Euclidean embedding space.  In the
standard basis let the embedding space metric have the form
$\eta_{AB}=diag(-++\ldots+-)$. We then consider the phase space with
coordinates $Y^A,P_A$ extended by ghost variables $c,\mu$ and their
associated ghost momenta $\pi,\rho$. The ghost degree is introduced by
prescribing $\gh{c}=\gh{\mu}=1$, $\gh{\pi}=\gh{\rho}=-1$, and
vanishing ghost degree to $Y,P$.

Let $\algA$ be the algebra of functions on the extended phase space
equipped with the Weyl star product determined by the basic
commutation relations:
\begin{equation}
\label{eq:alg-ads}
  \qcommut{P_A}{Y^B}=-\hbar\delta^A_B\,,\qquad \qcommut{\pi}{c}=-\hbar\,,\qquad \qcommut{\rho}{\mu}=-\hbar\,.
\end{equation}
The quantum BRST charge describing the scalar particle on AdS is given by
\begin{equation}
\label{eq:AdS-charge}
  \bar\Psi_0=c \,\ffrac{1}{2}P^A P_A+\mu Y^AP_A+2\mu c \pi\,,
\end{equation}
Note that the constraint $YP$ is defined up to adding a constant terms
without affecting the constraint algebra and hence the nilpotency of
$\bar\Psi_0$.  This quantum BRST charge and its generalization for HS
particles and tensionless strings on AdS are known in the literature
(see e.g.~\cite{Bonelli:2003zu}).  Note that if at the quantum level
one directly impose the constraints on wave functions one reproduces
the well-known approach from~\cite{Fronsdal:1979vb}.

Now we are going to put the constrained system to the target space.
To this end we follow the construction of Section~\bref{sec:general}
with $\algA$ and $\bar\Psi_0$ described above. More specifically, we
take $\algA$ to be the $*$-product algebra of functions in
$Y,P,c,\pi,\mu,\rho$ that are formal power series in $Y$ and
polynomials in the rest of the variables. Besides the $*$-product we
also equip $\algA$ with the BRST operator determined by $\bar\Psi_0$.  The associated
supermanifold $\manM$ is then equipped with the following
$Q$-structure:
\begin{equation}
Q\Psi=\bar\brst\Psi+\ffrac{1}{2\hbar}\qcommut{\Psi}{\Psi}\,,\qquad \quad
\bar\brst\phi=\ffrac{1}{\hbar}\qcommut{\bar\Psi_0}{\phi}\,.
\end{equation}

As a $Q$-manifold $\manX$ we take an odd tangent bundle over the
$AdS_\dmn$ space $\manX_0$ equipped with the De Rham differential.  As
before local coordinates on $\manX_0$ and the fibers of $\Pi T\manX_0$
are denoted by $x^\mu$ and $\theta^\mu$ so that $\derham=\theta^\mu
\dl{x^\mu}$.  We then consider a BRST field theory $(\manX,\derham,
\manM,Q)$. As we are going to see it describes the off-shell HS fields
on $AdS_\dmn$ space in terms of the $\dmn+1$-dimensional embedding
space.

\subsection{Linearization}\label{sec:ads-lin}
Similarly to the flat case considered in Section~\bref{sec:flat},
equation of motion for physical fields can be identified with the
master equation of the parent theory for a scalar particle on $AdS_d$.
A natural choice for a particular solution to
$\derham\Psi_0+\bar\brst\Psi_0+(2\hbar)^{-1}\qcommut{\Psi_0}{\Psi_0}=0$
is then to take $\Psi_0$ describing the parent system~\cite{\BGadS} for a
particle on AdS space. Namely, a particular solution reads as:
\begin{equation}
\label{eq:part-ads}
  \Psi_0=\theta^\mu e_\mu^A P_A+\theta^\mu \omega_{\mu B}^A (Y^B+V^B)P_A+\mu V^AP_A
\end{equation}
so that $\Psi_0+\bar\Psi_0$ indeed coincides with the quantum BRST charge
of the parent system from~\cite{\BGadS} (strictly speaking one should
also add $-\theta^\mu\bar p_\mu$, see Section~\bref{sec:general}). It
satisfies
$\derham\Psi_0+\bar\brst\Psi_0+(2\hbar)^{-1}\qcommut{\Psi_0}{\Psi_0}=0$
provided $e,\omega,V$ (identified respectively with a vielbein, $AdS$
connection, and a given section of the vector bundle with
$\dmn+1$-dimensional fiber) satisfy
\begin{equation}
\label{eq:EWV-comp}
\begin{aligned}
  d V^A+\omega^A_B V^B&=e^A,\qquad& d e^A+\omega_B^A e^B&=0\,, \\
  d\omega^A_B+\omega^A_C \omega^C_B&=0\,,\qquad& V^AV_A+l^2&=0\,,
\end{aligned}
\end{equation}
where $l$ is the radius of $AdS_\dmn$. In addition one has to require
$e$ to have a maximal rank, i.e. $\rank(e)=\dmn$.

Note that both $\bar\Psi_0$ and $\bar\Psi$ are frame-independent
provided the components of $e,\omega,V$ transform as those of a
vielbein, a connection, and a section respectively. In particular one
can alway chose the frame where $V^A={\rm const}^A$. In this case one
can redefine $\bar\Psi_0,\Psi_0$ according to $\bar\Psi_0\to
\bar\Psi_0+\mu V^AP_A$ and $\Psi_0\to \Psi_0-\mu V^AP_A$ in order to
completely distinguish the ``space-time'' and the ``target-space''
parts of the BRST differential.

 Linearizing $(\manX,\derham,\manM,Q)$ around $\Psi_0$ one gets the
following linear BRST differential
\begin{multline}
\label{eq:s0-ads}
  s_0\Psi=\Big[\nabla+\sigma-c P^A\dl{Y^A}
+
\mu(P_A\dl{P_A}-(Y^A+V^A)\dl{Y^A})
-\ffrac{1}{2}(P_A P^A)\dl{\pi}
-
\\
-(Y^A+V^A)P_A\dl{\rho}
-2\mu c\dl{c}
+ 2\mu \pi\dl{\pi}
-2c\pi \dl{\rho}
+\hbar^2(\ldots)
\Big]\Psi\,,
\end{multline}
where 
\begin{equation}
\nabla=  \derham-\omega_A^B Y^A\dl{Y^B}+\omega_A^B P_B\dl{P_A}\,,\qquad \sigma=-e^A\dl{Y^A}\,.
\end{equation}
and $(\ldots)$ denote some operator whose explicit form is not needed
for the moment.

This differential can be identified as that of the field theory
associated to the first-quantized system
$(\brst,\Gamma(\bundle\algA,\manX))$ with the string field $\Psi$ and
the BRST operator determined by $s_0\Psi=\brst\Psi$. Note that the
nilpotency of $\brst$ is guaranteed by the nilpotency of $s_0$ which
is in turn a consequence of the nilpotency of the original non-linear
BRST differential.  As we are going to see this first-quantized system
can be considered as the parent system (in the sense
of~\cite{\BGST,\BGadS}) describing off-shell HS fields on AdS space.

Similarly to the flat case the easiest way to see this is to reduce
the system described by~\eqref{eq:s0-ads} further. We do the necessary
reduction in two steps.  First we take as a degree homogeneity in
$\mu$ so that $\brst$ decomposes as $\brst=\brst_{-1}+\brst_{0}$ with
\begin{equation}
  \brst_{-1}=\mu(h-2c\dl{c}+2\pi\dl{\pi})\,,\qquad h=P^A\dl{P^A}-(Y^A+V^A)\dl{Y^A}\,.
\end{equation}
It was shown in~\cite{\BGadS} that the cohomology of operators of this
type in the space of formal power series in $Y$ is concentrated in
degree zero and hence is given by a subspace $\hat\cE\subset \algA$ of
elements annihilated by $h-2c\dl{c}+2\pi\dl{\pi}$.  The reduction is
then straightforward and gives the system $(\hat\brst,\Gamma(\bundle{\hat\cE},\manX))$ with
  \begin{equation}
  \hat\brst= \nabla+\sigma-c P^A\dl{Y^A}
-\ffrac{1}{2}(P_A P^A)\dl{\pi}-(Y^A+V^A)P_A\dl{\rho}- 2 c \pi \dl{\rho}+\hbar^2(\ldots)\,.
\end{equation}

The next step is to take as a degree homogeneity in $\pi$ and $\rho$. BRST operator $\hat\brst$
then decomposes as $\hat\brst=\hat\brst_{-1}+\hat\brst_0$ with
\begin{equation}
  \hat\brst_{-1}=-\ffrac{1}{2}(P_A P^A)\dl{\pi}-(Y^A+V^A)P_A\dl{\rho}+\hbar^2(\ldots)\,.
\end{equation}
It is not difficult to compute cohomology of $\hat\brst_{-1}$. Indeed,
the first term factors out traces (elements proportional to $P^2$) and
the second one allows to completely control the dependence of
representatives on $P^A V_A$ (i.e. on a component of $P$ parallel
along $V$). Let us introduce the following notations
\begin{equation}
  \begin{gathered}
    \Box=\dl{Y^A}\dl{Y_A},\qquad T=\dl{P^A}\dl{P_A},\qquad
    S=\dl{P^A}\dl{Y_A}, \\
 \bsd=(Y^A+V^A)\dl{P^A},\qquad
    \sd=P^A\dl{Y^A}.
\end{gathered}
\end{equation}
for some of generators of $sp(4)$ algebra identified in~\cite{\BGST,\BGadS}.
Leaving details of the proof to the Appendix~\bref{sec:ads-app} we
have
\begin{prop}\label{prop:ads}
  Cohomology of $\hat\brst_{-1}$ in the space of formal power series in
  $Y$ and polynomials in variables $P$ and ghosts is given by
  \begin{equation}
   \qquad H^0(\hat\brst_{-1},\hat\cE)\cong\cE\,,\qquad  H^n(\hat\brst_{-1},\cE)=0\quad n\neq 0\,,
  \end{equation}
  where $\cE\subset \ker \hat\brst_{-1}$ is the subspace of
  $\mu,\pi,\rho$-independent elements from $\algA$ satisfying
\begin{equation}
\label{eq:E-cond}
  T \phi=0\,,\qquad \bsd\phi=0\,,\qquad (h-2c\dl{c})\phi=0\,.
\end{equation}
\end{prop}
Note that the subspace~\eqref{eq:E-cond} is particularly convenient
for our purposes but, in general, one can use different identification
of cohomology as a subspace in $\ker\hat\brst_{-1}$. Using the
reduction technique from~\cite{\BGST} one then finds that the system
can be reduced to $(\tilde\brst,\Gamma(\bundle\cE,\manX))$. Because
$\brst_{-1}$-cohomology is concentrated in one degree only the
construction of $\tilde\brst$ is straightforward and gives:
\begin{equation}
  \tilde\brst=\nabla+\sigma-\Pj_{\cE}c\sd\,.
\end{equation}
Here, $\Pj_\cE$ denotes the projector to $\cE$ defined as follows: for
$\phi=\phi_0+\phi^\prime$ with $\phi_0\in\cE$ and $\phi^\prime\in \im
\hat\brst_{-1}$ one defines $\Pj_\cE\phi=\phi_0$ (see
Appendix~\bref{sec:ads-app} for more details).

The physical component of the string field is $\Psi^{(0)}=\theta^\mu
A_\mu(x;Y,A)+cF(x;Y,A)$, where we have introduced physical fields $A$
and $F$ identified with the 1-form and $0$-form on $\manX_0$
respectively. The equation of motion and gauge symmetries of the
associated field theory read as
\begin{gather}
  (\nabla +\sigma)A=0\,,\qquad  (\nabla +\sigma)F=-\Pj_\cE\sd A\,,\\
\delta_\lambda A=(\nabla +\sigma)\lambda\,,\qquad \delta_\lambda F=-\Pj_\cE\sd \lambda\,,
\end{gather}
where $\lambda$ is the gauge parameter.

In fact it can be also useful to consider this first-quantized system
without the trace constraint, i.e. with $\cE$ being the subspace of
$\pi,\rho$-independent elements from $\hat\cH$ satisfying
$\bsd\phi=(h-2c\dl{c})\phi=0$ only and the BRST operator given by
$\tilde\brst=\nabla+\sigma-c\sd$. The advantage is that no projectors
are needed but at the same time the system does not appear naturally
as a linearization of some consistent non-linear system. Reducing this
system further to cohomology of $c\sd$ one arrives at the
first-quantized system describing the linearized unfolded off-shell HS
theory equivalent to that from~\cite{Sagnotti:2005ns}.

The off-shell system $(\tilde\brst,\Gamma(\bundle\cE,\manX))$ can
obviously be restricted to the first-quantized system $(\brst_{\rm
  on-shell},\Gamma(\bundle{\cE_{\rm on-shell}},\manX))$, with
$\cE_{\rm on-shell}\subset \cE$ being the subspace of elements
satisfying the additional conditions $\Box\phi=S\phi=0$. Note that in
the restricted $\brst_{\rm on-shell}$ the projector is not anymore
needed because for $\phi$ satisfying $\phi\in\cE_{\rm on-shell}$ one
has $\sd\phi\in\cE_{\rm on-shell}$.  The resulting system then
explicitly coincides with the so-called intermediate
system~\cite{\BGadS} describing Fronsdal HS fields on AdS space. It
was shown in~~\cite{\BGadS} that the field theory associated to
$(\brst_{\rm on-shell},\Gamma(\bundle{\cE_{\rm on-shell}},\manX))$ is
indeed equivalent to the Fronsdal HS gauge theory via
elimination/addition of generalized auxiliary fields.  In particular,
reducing further to the cohomology of $c\sd$ one arrives~\cite{\BGadS}
at the familiar unfolded
formulation~\cite{Lopatin:1988hz,Vasiliev:2001wa}.

\subsection{Topological HS theory}
In Section~\bref{sec:flat} and \bref{sec:ads-lin} we have seen how the
linearized off-shell HS theory can be put on-shell in the case of flat
and AdS space respectively.  Putting the non-linear HS theory on-shell
implies constructing the interacting HS theory. At the level of
equations of motion this problem was solved by
M.~Vasiliev~\cite{Vasiliev:1990en,Vasiliev:2003ev} for HS fields on
AdS space.  Here, we do not discuss the entire Vasiliev construction.
Instead, we restrict to the sector of HS connections only and show
that the truncated system can be easily put on-shell giving the
``topological'' HS theory whose equations of motion are zero-curvature
equations for the HS connection taking values in the HS algebra.

To this end consider the linearized off-shell theory determined by the
BRST differential~\eqref{eq:s0-ads} but replace $\algA$ with the
algebra of polynomials in all the variables including $Y$ in contrast
to the formal power series in $Y$ considered in the previous section
(see also~\cite{\BGadS}). It was shown in~\cite{\BGadS} that in the
on-shell version, only ``one half'' of the fields survives if one
restricts to polynomials. Indeed, the HS connections are represented
by polynomials while the HS curvatures are represented by the formal
power series in $Y$-variables so that dropping formal power series
indeed corresponds to putting HS curvatures to zero.

In the space of polynomials it is legitimate to redefine $Y$-variables
according to $Y^A+V^A\to Y^A$. The expression for the BRST operator
then takes the form:
\begin{multline}
\label{eq:os-lin-ads}
\brst= \nabla-c P^A\dl{Y^A}+\mu(P^A\dl{P^A}-Y^A\dl{Y^A})
+\\
-\ffrac{1}{2}(P_A P^A)\dl{\pi}-(Y^A P_A)\dl{\rho}-2\mu(c\dl{c}-\pi\dl{\pi}) -2c
\pi \dl{\rho}+\hbar^2(\ldots)\,.
\end{multline}

If one restricts to polynomials the system can be put on-shell in a
different way.  Namely, let $c_0,b_0$ be the new Grassmann odd ghost
variables with $\gh{c_0}=1$ and $\gh{b_0}=-1$ and let
\begin{equation}
  {\brst^\prime}=\brst+c_0 Y^A\dl{P_A}-\ffrac{1}{2}(Y^A Y_A)\dl{b_0}+\text{ghost terms}
\end{equation}
where ``ghost terms'' are terms cubic in ghosts and derivatives with
respect to ghosts, needed to maintain nilpotency.  The adapted version
of the arguments given in~\cite{\BGadS} immediately show that the
first-quantized system determined by ${\brst^\prime}$ reduces to
$(\tilde\brst,\Gamma(\bundle{\algH\oplus\bar\algH},\manX))$ with
$\tilde\brst=\nabla$, $\algH$ being the space of totally traceless
polynomials in $Y,A$ described by the rectangular Young tableaux, and
$\bar\algH$ being the space of elements of the form $\bar\phi=\mu c c_0
\phi$ with $\phi\in\algH$.

Too see this let us reduce the theory to the cohomology of the
``fiber'' part ${\bar\brst}^\prime=\brst^\prime-\nabla$. First we
reduce to the cohomology of $Y^2\dl{b_0}+A^2\dl{\pi}+AY\dl{\rho}$,
which can be identified with $\rho,\pi,b_0$-independent totally
traceless elements, and then to the cohomology of the remaining
operators.  Because the remaining operators $Y^A\dl{P^A}$,
$P_A\dl{P_A}-Y^A\dl{Y^A}$, and $P^A\dl{Y^A}$ form the standard
representation of $sl(2)$ on polynomials the cohomology can indeed be
identified with the subspace $\algH\oplus \bar\algH$, i.e.
$sl(2)$-invariants in vanishing and top ghost degrees.  Because
physical fields are associated to elements of vanishing ghost number
only it follows that there are no physical fields associated to
$\bar\algH$ and therefore
$(\brst^\prime,\Gamma(\bundle{\algH\oplus\bar\algH},\manX))$ indeed
properly describes linearized HS connections on AdS.

One then observes that the linearized BRST differential determined by
${\brst^\prime}$ can also be identified as a linearization of some
non-linear BRST differential.  Indeed, let us extend the $\algA$ and
the Weyl $*$-product~\eqref{eq:alg-ads} with the additional ghost
variables $c_0,b_0$ such that $\qcommut{b_0}{c_0}=-\hbar$. One then
builds the BRST field theory $(\manX,\derham,\manM^\prime,Q^\prime)$,
where $\manM^\prime$ and $Q^\prime$ are, respectively, the associated
supermanifold and the $Q$-structure determined by
\begin{equation}
  Q\Psi^\prime=\bar\brst^\prime{\Psi^\prime}+
\ffrac{1}{2\hbar}\qcommut{\Psi^\prime}{\Psi^\prime}\,,\qquad
\bar\brst^\prime\phi=\ffrac{1}{\hbar}\qcommut{\bar\Psi^\prime_0}{\phi}\,,
\end{equation}
where
\begin{equation}
\label{eq:fiber-part}
  \bar\Psi^\prime_0=c \half P^A P_A+\mu Y^A P_A+c_0 \half Y^A Y_A 
+2\mu c \pi
-2\mu c_0 b_0
-c c_0 \rho\,.
\end{equation}
The BRST operator ${\brst^\prime}$ corresponds then to the following
choice of the particular solution $\Psi_0^\prime=\theta^\mu
\omega_{\mu B}^A Y^B P_A$ in the sense that
${\brst^\prime}=\derham+\bar\brst^\prime+\hbar^{-1}\qcommut{\Psi_0^\prime}{\cdot\,}$.

The ``fiber'' part $\bar\Psi_0^\prime$ given by~\eqref{eq:fiber-part}
is the standard quantum BRST charge for the system with three
constraints forming the $sl(2)$ algebra. In the adapted notations
$Y_1^{A}=Y^A$ and $Y_2^{A}=P^A$ the constraints take the form
$T_{ij}=Y_i^A Y_{jA}$.  Introducing the Lagrange multipliers
$\Lambda^{ij}=\Lambda^{ji}$, the extended Hamiltonian action for a
systems with these constraints takes the form\footnote{The author is
  grateful to Itzhak Bars for the illuminating discussion on this
  model and its relation to the two-time physics approach.}
\begin{equation}
\label{eq:sp2}
  S=\half \int dt ( Y_{Ai}\dl{t} Y^{Ai}+\Lambda^{ij}Y^A_i Y_{Aj})\,,
\end{equation}
familiar from~\cite{Bars:2000qm,Bars:2001um} (see
also~\cite{Engquist:2005yt} and references therein).  Note that $S$
possesses the gauge symmetry with the gauge algebra $sl(2)$ if one
identifies $\Lambda$ as a $sl(2)\cong sp(2)$-valued gauge field (in
the literature this gauge algebra is usually identified with $sp(2)$,
which is perhaps a more appropriate notation in this context). One
then concludes that the constrained system determined by the quantum
BRST charge $\Psi^\prime_0+\bar\Psi^\prime_0$ (as before one also
needs to add $-\theta^\mu \bar p_\mu$ to take $\derham$ into account)
is a version of the parent system constructed for the
model~\eqref{eq:sp2}.

To see explicitly the gauge field theory described by
$(\manX,\derham,\manM^\prime,Q^\prime)$ let us reduce the `fiber''
system $(\manM^\prime,Q^\prime)$ to the cohomology of the ``fiber''
part $\bar\brst^{\prime}$ of the BRST operator. It then follows that
BRST field theory can be reduced accordingly (see
Section~\bref{sec:reduction}). In its turn the cohomology of
$\bar\brst^\prime$ can be identified with the subspace
$\algH\oplus\bar\algH$ defined above.  The $*$-product determines the
associative product (also denoted by $*$) in $\algH$ by identifying
elements of $\algH$ with representatives of the BRST cohomology. One
then concludes that $\algH$ is precisely the HS algebra described
in~\cite{Eastwood:2002su,Vasiliev:2003ev}.  Note that $\algH$ can be
considered as the algebra of quantum observables of the system
determined by the quantum BRST charge $\bar\Psi_0^\prime$.

Expanding the theory around a particular solution
$\Psi_0^\prime=\theta^\mu \omega_{\mu B}^A Y^B P_A$ one finds the
equations of motion and gauge symmetries
\begin{equation}
  \derham A=\nabla A+\ffrac{1}{2\hbar}\qcommut{A}{A}\,,\qquad 
\delta_\lambda A=\derham \lambda +\ffrac{1}{\hbar}\qcommut{A}{\lambda}\,.
\end{equation}
Here $\lambda=\lambda(x;Y,P)$ is the $\algH$-valued gauge parameter
and $A=\theta^\mu A_\mu(x;Y,P)$ is the $\algH$-valued 1-form.  Note
that it follows form counting the ghost degree that $A_\mu$ is the
only physical (ghost-number-zero) field. The equation of motion is
then the standard zero curvature equation for $\algH$-valued
connection 1-form as it should be if one puts to zero the HS
curvatures in the full nonlinear system from~\cite{Vasiliev:2003ev}.

\section*{Acknowledgments}

The author wishes to thank the International Solvay Institutes for
hospitality. He is grateful to K.~Alkalaev, I.~Tipunin, M.~Vasiliev,
and especially to G.~Barnich for stimulating and useful discussions.  This
work is supported by the RFBR Grant 05-01-00996,
the RFBR--JSPS Grant 05-01-02934YaF\_a, and by the Grant
LSS-4401.2006.2.

\appendix

\section{Proof of the Proposition}\label{sec:ads-app} 
First we compute the cohomology of the ``classical'' part
$\hat\brst^{cl}_{-1}=\hat\brst_{-1}\big|_{\hbar=0}$ given explicitly
by
\begin{equation}
  \hat\brst^{cl}_{-1}=-\hhalf(P_A P^A)\dl{\pi}
-(Y^A+V^A)P_A\dl{\rho}\,.
\end{equation}
It is useful to divide the computation in two steps. First we reduce
to the cohomology of $P^2\dl{\pi}$, which can be identified with the
traceless $\pi$-independent elements.  In this subspace the reduced
operator acts as follows
\begin{multline}
\label{eq:PTbrst}
  \Pj_\T\hat\brst^{cl}_{-1}=-\Pj_\T(Y^A+V^A)P_A\dl{\rho}~=\\
=~-(Y^A+V^A)P_A\dl{\rho}+(P^A P_A)\frac{1}{(d+1)+2{P^A\dl{P_A}}}\bsd\dl{\rho}\,,
\end{multline}
where $\Pj_\T$ denotes the standard projection to the $T$-traceless
(i.e. annihilated by $T$) component.  

Any $\rho$-independent traceless element is a representative of a
cohomology class. Let us show that by using a coboundary freedom one
can always assume that a representative also satisfy $\bsd \phi=0$. To
this end let $\phi_0\in \cH$ satisfy $T\phi_0=0$ and
$(h-2c_0\dl{c_0})\phi_0=0$.  Let us assume also that we are in the
frame where $V^A=l\delta^A_d$ and use the notations $Y^{d}=lz, P^d=lw$
and $Y^a=y^a,P^a=p^a$.  Consider the following sequence
\begin{equation}
  \phi_{n+1}=\phi_n-(Y^A+V^A)P_A \frac{1}{(z+1)^2} M \bsd\phi_n-P^2 \frac{1}{(1+z)^2}\bsd N \bsd \phi_n\,,
\end{equation}
where $M,N$ are some coefficients depending on the dimension $d$, operators $p^a\dl{p^a}$ and $w\dl{w}$
counting the homogeneity in $p,w$. The requirement $T\phi_{n+1}=0$ fixes one coefficients in terms of another
if one assumes $T\phi_n=0$. In particular, $\phi_{n+1}-\phi_n$ is in the image of~\eqref{eq:PTbrst}.

Let us introduce the filtration $\cH=\cE_0 \supset\cE_1\supset
\cE_2\supset\ldots \supset \cE_n \supset \ldots$ where $\cE_n$
consists of elements of degree greater or equal than $n$, with the
degree introduced according to $\deg{p}=\deg{y}=\deg{z}=2$ and
$\deg{w}=1$. Using the remaining freedom in the coefficients one can
always achieve that $\bsd \phi_{n+1}\in \cE_{n+1}$ provided $\bsd
\phi_n \in \cE_n$.

To show that cohomology of $\hat\brst_{-1}=\hat\brst_{-1}^{cl}+\hbar^2(\ldots)$
is the same we need an explicit form of the $\hbar^2$-term:
\begin{equation}
\hat\brst_{-1}=\hat\brst_{-1}^{cl}-\ffrac{\hbar^2}{8}(\Box\dl{\pi}-2S \dl{\rho})\,.
\end{equation}
It then follows that any $\rho,\pi$-independent element from
$\cH$ is a cocycle of $\hat\brst_{-1}$ as well. By iterating the argument above order by order in $\hbar$
one shows that by adding a coboundary one can always assume a representative to be annihilated by $\bsd$ and $T$.
The decomposition of $\cH$ then reads
\begin{equation}
\cH=\cE\oplus \im \hat\brst_{-1}\oplus\cF\,,
\end{equation}
where $\cF$ denote the complementary subspace (note that this subspace
contains $\pi,\rho$-dependent elements only). This determines the
explicit form of the projector $\Pj_\cE$ which we need in the main
text.

\providecommand{\href}[2]{#2}\begingroup\raggedright\endgroup


\begin{thebibliography}{10}

\bibitem{Vasiliev:2003ev}
M.~A. Vasiliev, ``{N}onlinear equations for symmetric massless higher spin
  fields in ({A})d{S}(d),'' {\em Phys. Lett.} {\bf B567} (2003) 139--151,
\href{http://www.arXiv.org/abs/hep-th/0304049}{{\tt hep-th/0304049}}.

\bibitem{Bekaert:2005vh}
X.~Bekaert, S.~Cnockaert, C.~Iazeolla, and M.~A. Vasiliev, ``Nonlinear higher
  spin theories in various dimensions,''
\href{http://www.arXiv.org/abs/hep-th/0503128}{{\tt hep-th/0503128}}.

\bibitem{Sagnotti:2005ns}
A.~Sagnotti, E.~Sezgin, and P.~Sundell, ``On higher spins with a strong
  {Sp(2,R)} condition,''
\href{http://www.arXiv.org/abs/hep-th/0501156}{{\tt hep-th/0501156}}.

\bibitem{Vasiliev:2005zu}
M.~A. Vasiliev, ``Actions, charges and off-shell fields in the unfolded
  dynamics approach,'' {\em Int. J. Geom. Meth. Mod. Phys.} {\bf 3} (2006)
  37--80,
\href{http://www.arXiv.org/abs/hep-th/0504090}{{\tt hep-th/0504090}}.

\bibitem{Batalin:1981jr}
I.~A. Batalin and G.~A. Vilkovisky, ``Gauge algebra and quantization,'' {\em
  Phys. Lett.} {\bf B102} (1981)
27--31.

\bibitem{Batalin:1983jr}
I.~A. Batalin and G.~A. Vilkovisky, ``Quantization of gauge theories with
  linearly dependent generators,'' {\em Phys. Rev.} {\bf D28} (1983)
2567--2582.

\bibitem{Barnich:2006pc}
G.~Barnich and M.~Grigoriev, ``Parent form for higher spin fields on anti-de
  {S}itter space,''
\href{http://www.arXiv.org/abs/hep-th/0602166}{{\tt hep-th/0602166}}.

\bibitem{Fradkin:1975cq}
E.~S. Fradkin and G.~A. Vilkovisky, ``Quantization of relativistic systems with
  constraints,'' {\em Phys. Lett.} {\bf B55} (1975)
224.

\bibitem{Batalin:1977pb}
I.~A. Batalin and G.~A. Vilkovisky, ``Relativistic {S} matrix of dynamical
  systems with boson and fermion constraints,'' {\em Phys. Lett.} {\bf B69}
  (1977)
309--312.

\bibitem{Bars:2000qm}
I.~Bars, ``Survey of two-time physics,'' {\em Class. Quant. Grav.} {\bf 18}
  (2001) 3113--3130,
\href{http://www.arXiv.org/abs/hep-th/0008164}{{\tt hep-th/0008164}}.

\bibitem{Segal:2000ke}
A.~Y. Segal, ``Point particle: Symmetric tensors interaction,''
\href{http://www.arXiv.org/abs/hep-th/0008105}{{\tt hep-th/0008105}}.

\bibitem{Segal:2002gd}
A.~Y. Segal, ``Conformal higher spin theory,'' {\em Nucl. Phys.} {\bf B664}
  (2003) 59--130,
\href{http://www.arXiv.org/abs/hep-th/0207212}{{\tt hep-th/0207212}}.

\bibitem{Engquist:2005yt}
J.~Engquist and P.~Sundell, ``Brane partons and singleton strings,''
\href{http://www.arXiv.org/abs/hep-th/0508124}{{\tt hep-th/0508124}}.

\bibitem{Hata:1993pp}
H.~Hata, ``'{T}heory of {T}heories' approach to string theory,'' {\em Phys.
  Rev.} {\bf D50} (1994) 4079--4087,
\href{http://www.arXiv.org/abs/hep-th/9308001}{{\tt hep-th/9308001}}.

\bibitem{Barnich:2005ru}
G.~Barnich and M.~Grigoriev, ``{BRST} extension of the non-linear unfolded
  formalism,''
\href{http://www.arXiv.org/abs/hep-th/0504119}{{\tt hep-th/0504119}}.

\bibitem{Vasiliev:1990en}
M.~A. Vasiliev, ``Consistent equation for interacting gauge fields of all spins
  in (3+1)-dimensions,'' {\em Phys. Lett.} {\bf B243} (1990)
378--382.

\bibitem{Vasiliev:1992av}
M.~A. Vasiliev, ``More on equations of motion for interacting massless fields
  of all spins in (3+1)-dimensions,'' {\em Phys. Lett.} {\bf B285} (1992)
225--234.

\bibitem{Alexandrov:1997kv}
M.~Alexandrov, M.~Kontsevich, A.~Schwartz, and O.~Zaboronsky, ``The geometry of
  the master equation and topological quantum field theory,'' {\em Int. J. Mod.
  Phys.} {\bf A12} (1997) 1405--1430,
\href{http://www.arXiv.org/abs/hep-th/9502010}{{\tt hep-th/9502010}}.

\bibitem{Fedosov:1994}
B.~Fedosov, ``A simple geometrical construction of deformation quantization,''
  {\em J. Diff. Geom.} {\bf 40} (1994) 213--238.

\bibitem{Fedosov-book}
B.~Fedosov, ``Deformation quantization and index theory,''. Berlin, Germany:
  Akademie-Verl. (1996) 325 p. (Mathematical topics: 9).

\bibitem{Bordemann:1997er}
M.~Bordemann, N.~Neumaier, and S.~Waldmann, ``Homogeneous {F}edosov star
  products on cotangent bundles {II}: {GNS} representations, the {WKB}
  expansion, and applications,''
\href{http://www.arXiv.org/abs/q-alg/9711016}{{\tt q-alg/9711016}}.

\bibitem{Batalin:1989mb}
I.~A. Batalin, E.~S. Fradkin, and T.~E. Fradkina, ``Generalized canonical
  quantization of dynamical systems with constraints and curved phase space,''
  {\em Nucl. Phys.} {\bf B332} (1990)
723.

\bibitem{Batalin:2001je}
I.~A. Batalin, M.~A. Grigoriev, and S.~L. Lyakhovich, ``Star product for second
  class constraint systems from a {BRST} theory,'' {\em Theor. Math. Phys.}
  {\bf 128} (2001) 1109--1139,
\href{http://www.arXiv.org/abs/hep-th/0101089}{{\tt hep-th/0101089}}.

\bibitem{Batalin:2005df}
I.~Batalin, M.~Grigoriev, and S.~Lyakhovich, ``Non-{A}belian conversion and
  quantization of non-scalar second-class constraints,'' {\em J. Math. Phys.}
  {\bf 46} (2005) 072301,
\href{http://www.arXiv.org/abs/hep-th/0501097}{{\tt hep-th/0501097}}.

\bibitem{Barnich:2004cr}
G.~Barnich, M.~Grigoriev, A.~Semikhatov, and I.~Tipunin, ``Parent field theory
  and unfolding in {BRST} first-quantized terms,'' {\em Commun. Math. Phys.}
  {\bf 260} (2005) 147--181,
\href{http://www.arXiv.org/abs/hep-th/0406192}{{\tt hep-th/0406192}}.

\bibitem{Segal:2001di}
A.~Y. Segal, ``Point particle in general background fields vs. free gauge
  theories of traceless symmetric tensors,'' {\em Int. J. Mod. Phys.} {\bf A18}
  (2003) 4999--5021,
\href{http://www.arXiv.org/abs/hep-th/0110056}{{\tt hep-th/0110056}}.

\bibitem{Barnich:2003wj}
G.~Barnich and M.~Grigoriev, ``Hamiltonian {BRST} and {B}atalin-{V}ilkovisky
  formalisms for second quantization of gauge theories,'' {\em Commun. Math.
  Phys.} {\bf 254} (2005) 581--601,
\href{http://www.arXiv.org/abs/hep-th/0310083}{{\tt hep-th/0310083}}.

\bibitem{Gaberdiel:1997ia}
M.~R. Gaberdiel and B.~Zwiebach, ``Tensor constructions of open string theories
  {I}: Foundations,'' {\em Nucl. Phys.} {\bf B505} (1997) 569--624,
\href{http://www.arXiv.org/abs/hep-th/9705038}{{\tt hep-th/9705038}}.

\bibitem{Fradkin:1985am}
E.~S. Fradkin and A.~A. Tseytlin, ``Conformal supergravity,'' {\em Phys. Rept.}
  {\bf 119} (1985)
233--362.

\bibitem{Fradkin:1989md}
E.~S. Fradkin and V.~Y. Linetsky, ``Cubic interaction in conformal theory of
  integer higher spin fields in four-dimensional space-time,'' {\em Phys.
  Lett.} {\bf B231} (1989)
97.

\bibitem{Fronsdal:1978rb}
C.~Fronsdal, ``Massless fields with integer spin,'' {\em Phys. Rev.} {\bf D18}
  (1978)
3624.

\bibitem{Fronsdal:1979vb}
C.~Fronsdal, ``Singletons and massless, integral spin fields on de {S}itter
  space (elementary particles in a curved space {VII}),'' {\em Phys. Rev.} {\bf
  D20} (1979)
848--856.

\bibitem{Mikhailov:2002bp}
A.~Mikhailov, ``Notes on higher spin symmetries,''
\href{http://www.arXiv.org/abs/hep-th/0201019}{{\tt hep-th/0201019}}.

\bibitem{Barnich:2000zw}
G.~Barnich, F.~Brandt, and M.~Henneaux, ``Local {BRST} cohomology in gauge
  theories,'' {\em Phys. Rept.} {\bf 338} (2000) 439--569,
\href{http://www.arXiv.org/abs/hep-th/0002245}{{\tt hep-th/0002245}}.

\bibitem{Lada:1993wc}
T.~Lada and J.~Stasheff, ``Introduction to {SH} {L}ie algebras for
  physicists,'' {\em Int. J. Theor. Phys.} {\bf 32} (1993) 1087--1104,
\href{http://www.arXiv.org/abs/hep-th/9209099}{{\tt hep-th/9209099}}.

\bibitem{Stasheff:1963}
J.~Stasheff, ``On the homotopy associativity of h-spaces, i, ii,'' {\em Trans.
  Amer. Math. Soc.} {\bf 108} (1963) 275--293.

\bibitem{Lyakhovich:2004xd}
S.~L. Lyakhovich and A.~A. Sharapov, ``{BRST} theory without hamiltonian and
  lagrangian,'' {\em JHEP} {\bf 03} (2005) 011,
\href{http://www.arXiv.org/abs/hep-th/0411247}{{\tt hep-th/0411247}}.

\bibitem{Schwarz:1992gs}
A.~Schwarz, ``Semiclassical approximation in {B}atalin-{V}ilkovisky
  formalism,'' {\em Commun. Math. Phys.} {\bf 158} (1993) 373--396,
\href{http://www.arXiv.org/abs/hep-th/9210115}{{\tt hep-th/9210115}}.

\bibitem{Sullivan}
D.~Sullivan, ``Infinitesimal computations in topology,'' {\em Inst. des Haut
  Etud. Sci. Pub. Math.} {\bf 47} (1977) 269.

\bibitem{Cattaneo:1999fm}
A.~S. Cattaneo and G.~Felder, ``A path integral approach to the {K}ontsevich
  quantization formula,'' {\em Commun. Math. Phys.} {\bf 212} (2000) 591--611,
\href{http://www.arXiv.org/abs/math.qa/9902090}{{\tt math.qa/9902090}}.

\bibitem{Batalin:2001fh}
I.~Batalin and R.~Marnelius, ``Generalized {P}oisson sigma models,'' {\em Phys.
  Lett.} {\bf B512} (2001) 225--229,
\href{http://www.arXiv.org/abs/hep-th/0105190}{{\tt hep-th/0105190}}.

\bibitem{Cattaneo:2001ys}
A.~S. Cattaneo and G.~Felder, ``On the {AKSZ} formulation of the {P}oisson
  sigma model,'' {\em Lett. Math. Phys.} {\bf 56} (2001) 163--179,
\href{http://www.arXiv.org/abs/math.qa/0102108}{{\tt math.qa/0102108}}.

\bibitem{Park:2000au}
J.-S. Park, ``Topological open p-branes,''
\href{http://www.arXiv.org/abs/hep-th/0012141}{{\tt hep-th/0012141}}.

\bibitem{Roytenberg:2002nu}
D.~Roytenberg, ``On the structure of graded symplectic supermanifolds and
  courant algebroids,''
\href{http://www.arXiv.org/abs/math.sg/0203110}{{\tt math.sg/0203110}}.

\bibitem{Kazinski:2005eb}
P.~O. Kazinski, S.~L. Lyakhovich, and A.~A. Sharapov, ``Lagrange structure and
  quantization,'' {\em JHEP} {\bf 07} (2005) 076,
\href{http://www.arXiv.org/abs/hep-th/0506093}{{\tt hep-th/0506093}}.

\bibitem{Ikeda:2006wd}
N.~Ikeda, ``{D}eformation of {B}atalin-{V}ilkovisky {S}tructures,''
\href{http://www.arXiv.org/abs/math.sg/0604157}{{\tt math.sg/0604157}}.

\bibitem{Grigoriev:1999qz}
M.~A. Grigoriev and P.~H. Damgaard, ``Superfield {BRST} charge and the master
  action,'' {\em Phys. Lett.} {\bf B474} (2000) 323--330,
\href{http://www.arXiv.org/abs/hep-th/9911092}{{\tt hep-th/9911092}}.

\bibitem{Vasiliev:1994gr}
M.~A. Vasiliev, ``Unfolded representation for relativistic equations in (2+1)
  anti-{D}e {S}itter space,'' {\em Class. Quant. Grav.} {\bf 11} (1994)
649--664.

\bibitem{Vasiliev:1988xc}
M.~A. Vasiliev, ``Equations of motion of interacting massless fields of all
  spins as a free differential algebra,'' {\em Phys. Lett.} {\bf B209} (1988)
491--497.

\bibitem{Dresse:1990dj}
A.~Dresse, P.~Gr\'egoire, and M.~Henneaux, ``Path integral equivalence between
  the extended and nonextended {H}amiltonian formalisms,'' {\em Phys. Lett.}
  {\bf B245} (1990) 192.

\bibitem{Vasiliev:2001zy}
M.~A. Vasiliev, ``Conformal higher spin symmetries of {4D} massless
  supermultiplets and {osp(L,2M)} invariant equations in generalized
  (super)space,'' {\em Phys. Rev.} {\bf D66} (2002) 066006,
\href{http://www.arXiv.org/abs/hep-th/0106149}{{\tt hep-th/0106149}}.

\bibitem{Gelfond:2003vh}
O.~A. Gelfond and M.~A. Vasiliev, ``Higher rank conformal fields in the
  {Sp(2M)} symmetric generalized space-time,''
\href{http://www.arXiv.org/abs/hep-th/0304020}{{\tt hep-th/0304020}}.

\bibitem{Grigoriev:2000rn}
M.~A. Grigoriev and S.~L. Lyakhovich, ``Fedosov deformation quantization as a
  {BRST} theory,'' {\em Commun. Math. Phys.} {\bf 218} (2001) 437--457,
\href{http://www.arXiv.org/abs/hep-th/0003114}{{\tt hep-th/0003114}}.

\bibitem{Bonelli:2003zu}
G.~Bonelli, ``On the covariant quantization of tensionless bosonic strings in
  {AdS} spacetime,'' {\em JHEP} {\bf 11} (2003) 028,
\href{http://www.arXiv.org/abs/hep-th/0309222}{{\tt hep-th/0309222}}.

\bibitem{Lopatin:1988hz}
V.~E. Lopatin and M.~A. Vasiliev, ``Free massless bosonic fields of arbitrary
  spin in d- dimensional de {S}itter space,'' {\em Mod. Phys. Lett.} {\bf A3}
  (1988)
257.

\bibitem{Vasiliev:2001wa}
M.~A. Vasiliev, ``Cubic interactions of bosonic higher spin gauge fields in
  {A}d{S}(5),'' {\em Nucl. Phys.} {\bf B616} (2001) 106--162,
\href{http://www.arXiv.org/abs/hep-th/0106200}{{\tt hep-th/0106200}}.

\bibitem{Bars:2001um}
I.~Bars and C.~Deliduman, ``High spin gauge fields and two-time physics,'' {\em
  Phys. Rev.} {\bf D64} (2001) 045004,
\href{http://www.arXiv.org/abs/hep-th/0103042}{{\tt hep-th/0103042}}.

\bibitem{Eastwood:2002su}
M.~G. Eastwood, ``Higher symmetries of the {L}aplacian,''
\href{http://www.arXiv.org/abs/hep-th/0206233}{{\tt hep-th/0206233}}.

\end{thebibliography}
\end{document}